\documentclass[a4paper,fleqn,usenatbib]{mnras}
\usepackage{subcaption}

\usepackage[normalem]{ulem}

\usepackage{newtxtext,newtxmath}
\usepackage{siunitx}
\usepackage[T1]{fontenc}
\usepackage{ae,aecompl}

\usepackage{graphicx}	% Including figure files
\usepackage{amsmath}	% Advanced maths commands
\usepackage{pdflscape}
\usepackage{hyperref}
\usepackage[dvipsnames]{xcolor}

\defcitealias{2022MNRAS.517..484N}{I}
\defcitealias{2023AN....34430006G}{II}

\title[NGC\,2506 with NIRCam@JWST]{Photometry and astrometry with \textit{JWST} -- III. 
A NIRCam-Gaia~DR3 analysis of the open cluster NGC\,2506}

\author[D.\ Nardiello]{D.\ Nardiello$^{1,2}$\thanks{E-mail: domenico.nardiello@inaf.it}, 
L.\ R.\ Bedin$^{1}$, 
M.\ Griggio$^{3,1}$,
M.\ Salaris$^{4,5}$,
M.\ Scalco$^{3,1}$,
S.\ Cassisi$^{5,6}$\\
$^{1}$Istituto Nazionale di Astrofisica - Osservatorio Astronomico di Padova, Vicolo dell'Osservatorio 5, IT-35122, Padova, Italy \\
$^{2}$Aix Marseille Univ, CNRS, CNES, LAM, Marseille, France \\
$^{3}$Dipartimento di Fisica, Universit\`a di Ferrara, Via Giuseppe Saragat 1, I-44122, Ferrara, Italy\\
$^{4}$Astrophysics Research Institute, Liverpool John Moores University, 146 Brownlow Hill, Liverpool L3 5RF, UK \\
$^{5}$Istituto Nazionale di Astrofisica - Osservatorio Astronomico di Abruzzo, Via M. Maggini, I-64100, Teramo, Italy \\
$^{6}$INFN - Sezione di Pisa, Largo Pontecorvo 3, I-56127 Pisa, Italy  \\
}

\date{Accepted 2023 August 5. Received 2023 July 31; in original form 2023 July 6}

\pubyear{2023}
\begin{document}
\label{firstpage}
\pagerange{\pageref{firstpage}--\pageref{lastpage}}
\maketitle

% Abstract of the paper
\begin{abstract}
In the third paper of this series aimed at developing the tools for
analysing resolved stellar populations using the cameras on board of
the \textit{James Webb Space Telescope (JWST)}, we present a detailed
multi-band study of the 2\,Gyr Galactic open cluster NGC\,2506.
We employ public calibration data-sets collected in multiple filters
to: 
\textit{(i)} derive improved effective Point Spread Functions (ePSFs)
for ten NIRCam filters;
\textit{(ii)} extract high-precision photometry and astrometry for
stars in the cluster, approaching the main-sequence (MS) lower mass of
$\sim 0.1~M_{\odot}$; and
\textit{(iii)} take advantage of the synergy between {\it JWST} and
\textit{Gaia}\,DR3 to perform a comprehensive analysis of the
cluster's global and local properties.
We derived a MS binary fraction of $\sim$57.5\,\%, extending the
\textit{Gaia} limit ($\sim 0.8~M_{\odot}$) to lower masses ($\sim
0.4~M_{\odot}$) with {\it JWST}.
We conducted a study on the mass functions (MFs) of NGC\,2506, mapping
the mass segregation with Gaia data, and extending MFs to lower masses
with the \textit{JWST} field. We also combined information on the
derived MFs to infer an estimate of the cluster present-day total
mass.
Lastly, we investigated the presence of white dwarfs (WDs) and
identified a strong candidate. However, to firmly establish its
cluster membership, as well as that of four other WD candidates and of
the majority of faint low-mass MS stars, further {\it JWST} equally
deep observations will be required.
We make publicly available catalogues, atlases, and the improved
ePSFs.

\end{abstract}

% Select between one and six entries from the list of approved keywords.
% Don't make up new ones.

%%%%%%%%%%%%%%%%%%%%%%%%%%%%%%%%%%%%%%%%%%%%%%%%%%

%%%%%%%%%%%%%%%%% BODY OF PAPER %%%%%%%%%%%%%%%%%%
\section{Introduction}

When we observe a Galactic stellar cluster, whether it is open or
globular, it is like looking at a snapshot of stars with different
masses but approximately the same initial chemical composition, age,
and distance. By studying stellar clusters with varying chemical
compositions and ages ranging from a few tens of Myr up to $\sim
10$--13~Gyr, we can gain insights into the processes involved in the
formation and evolution of stars with different mass and
metallicity. Optical and UV studies using data from the {\it Hubble
  Space Telescope} ({\it HST}) have led to the discovery of several
phenomena that were previously unknown, such as, for example, the
existence of multiple stellar populations in globular clusters (see,
e.g., \citealt{2004ApJ...605L.125B,
  2007ApJ...661L..53P,2015AJ....149...91P}), and the unusual shape of
some stellar clusters' white dwarf cooling sequences (e.g.,
\citealt{2008ApJ...678.1279B,2013ApJ...769L..32B}).
However, with few exceptions (see, e.g.,
\citealt{2005AJ....130..626K,2008AJ....135.2141R,2016ApJ...817...48D}),
observations of open and globular clusters have been primarily focused
on main-sequence (MS) stars with masses $\gtrsim 0.1~M_{\odot}$, and
it has been challenging to study stars near the hydrogen burning limit
(HBL) and the brown dwarf sequence. In this regard, infrared (IR)
photometry plays a crucial role in providing information about
low-mass (pre-)MS stars ($M \lesssim$0.1--0.2~$M_{\odot}$) and brown
dwarfs in stellar clusters (\citealt{2023MNRAS.521L..39N}).

Since July 2022, the {\it James Webb Space Telescope} ({\it JWST},
\citealt{2023PASP..135f8001G}) has been acquiring a vast amount of IR
data through its cameras, revolutionising our understanding of the
Universe.  Public Director's Discretionary-Early Release Science
(DD-ERS) and Calibration {\it JWST} data constitute a treasure for
improving data reduction techniques and at the same time yield
scientifically significant results (see, e.g., \citealt[hereafter
  Paper\,I and II, respectively]{2022MNRAS.517..484N,
  2023AN....34430006G}).  Currently, non-proprietary data in the
archive encompass observations of high-redshift galaxies
(e.g., \citealt{2022ApJ...940L..14N}) and galaxy clusters
(e.g., \citealt{2023arXiv230102179P}), resolved close dwarf galaxies and
portions of the Large Magellanic Cloud (LMC, e.g.,
Paper~\citetalias{2023AN....34430006G};
\citealt{2023arXiv230300009L}), Galactic stellar clusters
(e.g., Paper~\citetalias{2022MNRAS.517..484N};
\citealt{2023arXiv230406026Z}), individual stars and exoplanets
(e.g., \citealt{2023Natur.614..670F}), and objects of the Solar System
(e.g., \citealt{2022DPS....5430607D}).

In this study, we took advantage of the publicly available calibration
observations to conduct an IR multi-band investigation of the lower MS
stars in the open cluster NGC\,2506. The data were collected with the
{\it JWST} Near Infrared Camera (NIRCam,
\citealt{2023PASP..135b8001R}) as part of the calibration programme
CAL-1538 (PI: Gordon). The primary objective of this programme is to
acquire observations of G dwarf stars for the flux calibration of {\it
  JWST} filters. Additionally, we utilised data from the CAL-1476
programme (PI: Boyer) targeting stars in the LMC to derive effective
Point Spread Functions for nearly all available NIRCam filters,
spanning a wavelength range from $\sim 0.7~\mu$m to $\sim 4.5~\mu$m.

The open cluster NGC\,2506 is particularly interesting since it
belongs to the (small) sample of metal-poor ([Fe/H]$\lesssim -0.1$),
old-age ($\gtrsim 1$~Gyr) open clusters located at the Galactic
anti-centre, at $\sim 3$~kpc from the Sun
(\citealt{1981ApJ...243..841M,2020A&A...633A..99C}); similar clusters
in this category include NGC\,2420 and NGC\,2243
(\citealt{2005AJ....129..872A,2006AJ....131..461A}). NGC\,2506 has
been the subject of several studies to determine its age and
metallicity
(\citealt{2004A&A...422..951C,2011MNRAS.416.1092M,2016AJ....152..192A,2020MNRAS.499.1312K}),
analyse the evolution of surface lithium abundances
(\citealt{2018AJ....155..138A}), investigate its structural parameters
(\citealt{2013MNRAS.432.1672L, 2019MNRAS.490.1383R,
  2020ApJ...894...48G}), and identify binary systems and blue
straggler stars (\citealt{2007A&A...465..965A,
  2022MNRAS.516.5318P}). There have been some discrepancies among
these studies regarding the age and metallicity of the cluster
members, although all analyses provide an age range between 1 and 3
Gyr and classify the cluster members as highly metal-deficient, with
an upper limit of [Fe/H] around $-0.2$
(\citealt{2016A&A...585A.150N}). From a dynamical perspective,
NGC\,2506 is of significant interest. Several studies suggest that the
cluster is dynamically relaxed, displaying clear evidence of mass
segregation, evaporation of low-mass stars, and even hints of tidal
tails
(\citealt{2013MNRAS.432.1672L,2019MNRAS.490.1383R,2020ApJ...894...48G}).

In this work, we investigate all these aspects of this peculiar open
cluster, by combining {\it JWST} photometry and astrometry with
Gaia~DR3 (\citealt{2021A&A...649A...1G}) and ground-based
data. Section~\ref{sec:obs} reports the technical part of this work;
it includes a description of the adopted space and ground-based
observations, the description of the procedure to derive the effective
point spread functions, and the data
reduction. Section~\ref{sec:astrophoto} describes the photometric and
astrometric properties of the stars, while Sect.~\ref{sec:radprof}
includes the derivation of the radial stellar density profile of the
cluster and its structural parameters.  Section~\ref{sec:binmf}
reports in detail the analysis of the MS binary fraction and of the
mass functions extracted from both {\it JWST} and Gaia data, while
Sect.~\ref{sec:wd} discusses the candidate white dwarfs we found. A
summary is reported in Sect.~\ref{sec:sum}.

\section{Observations and data reduction}
\label{sec:obs}
In this section, we report a description of the {\it JWST} data used
in this work and the procedures adopted to derive effective point
spread functions (ePSFs) and the astro-photometric catalogues of
NGC\,2506. We also describe how we obtained the catalogues from both
ground-based data-sets and selections of the Gaia~DR3.

\subsection{Large Magellanic cloud {\it JWST} data-set}
\label{sec:lmc}

\begin{figure*}
\includegraphics[bb=15 119 559 575,width=0.9\textwidth]{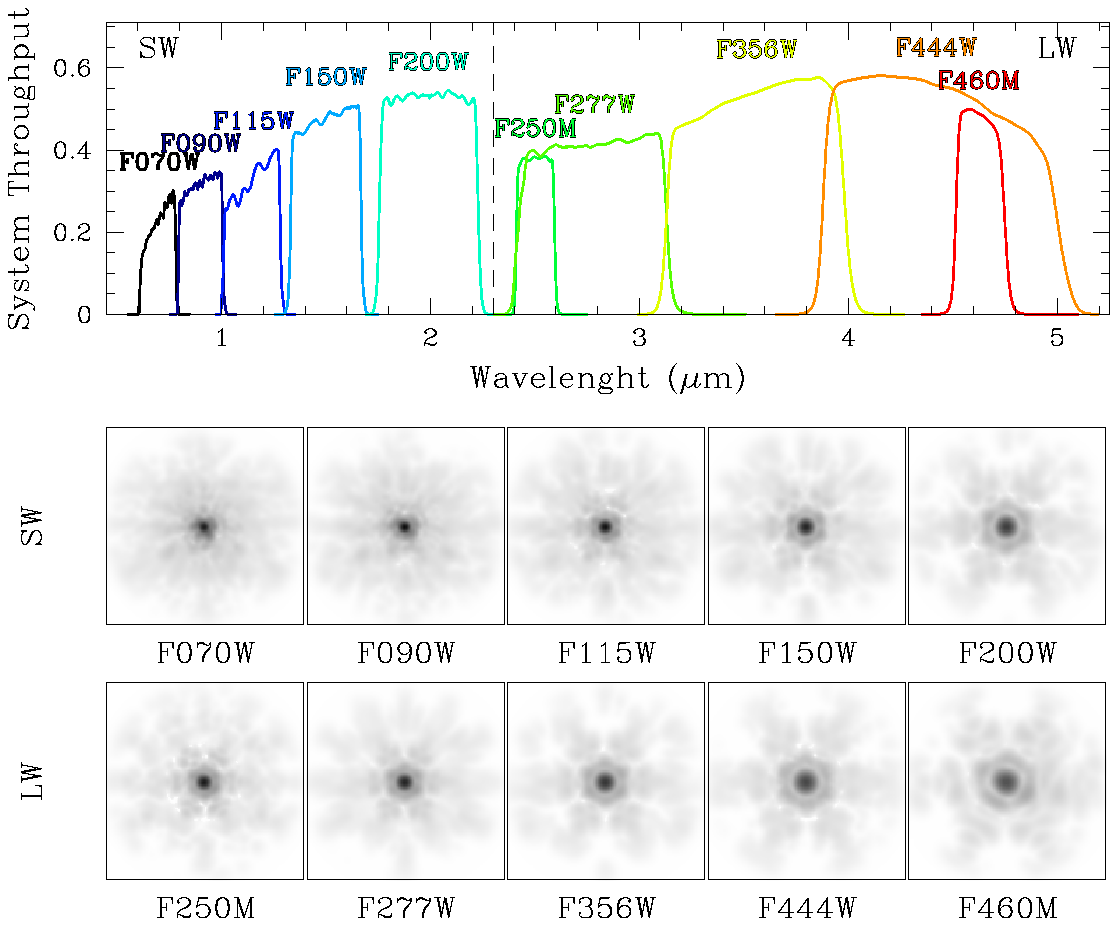}
\caption{ The {\it JWST} filters and ePSFs used in this work. Top
  panel shows the NIRCam + {\it JWST} total throughputs for the ten
  filters adopted in this work. The lower panels show the ePSFs in the
  central region of the detector \texttt{A1} (SW channel, middle row)
  and \texttt{ALONG} (LW channel, bottom row) for all the filters used
  in this work.  \label{fig:1}}
\end{figure*}

We employed the NIRCam@{\it JWST} data of the LMC collected during the
Calibration Programme CAL-1476 (PI: Boyer) to derive the ePSFs in 10
different filters. Specifically, we used images collected with the
Short Wavelength (SW) channel in F070W, F090W, F115W, F150W, and F200W
filters, and with the Long Wavelength (LW) channel in F250M, F277W,
F356W, F444W, and F460M filters. As detailed in Paper
\citetalias{2023AN....34430006G}, nine pointings are available for
each filter. Observations in F277W and F356W filters were conducted
using the \texttt{BRIGHT1} readout pattern and an effective exposure
time $t_{\rm exp}=96.631$~s (corresponding to 5 groups of 1 averaged
frame). For the F460M filter, the \texttt{BRIGHT2} readout pattern and
$t_{\rm exp}=85.894$~s (4 groups of 2 averaged frames) were employed.
Images in all the other filters were taken using the \texttt{RAPID}
readout patter, with 2 groups containing 1 frame ( $t_{\rm
  exp}=21.474$~s). The top panel of Fig.~\ref{fig:1} displays the
Total System Throughput for each filter.

For our analysis, we utilised the \texttt{\_cal} images created by the
Stage 2 pipeline
\texttt{calwebb\_image2}\footnote{\url{https://jwst-pipeline.readthedocs.io/}};
we converted the value of each pixel from MJy/sr into counts by using
the header keywords \texttt{PHOTMJSR} and
\texttt{XPOSURE}. Additionally, we flagged bad and saturated pixels by
using the Data Quality image included in the \texttt{\_cal} data cube.

\begin{figure*}
\includegraphics[bb=23 223 568 553,width=0.9\textwidth]{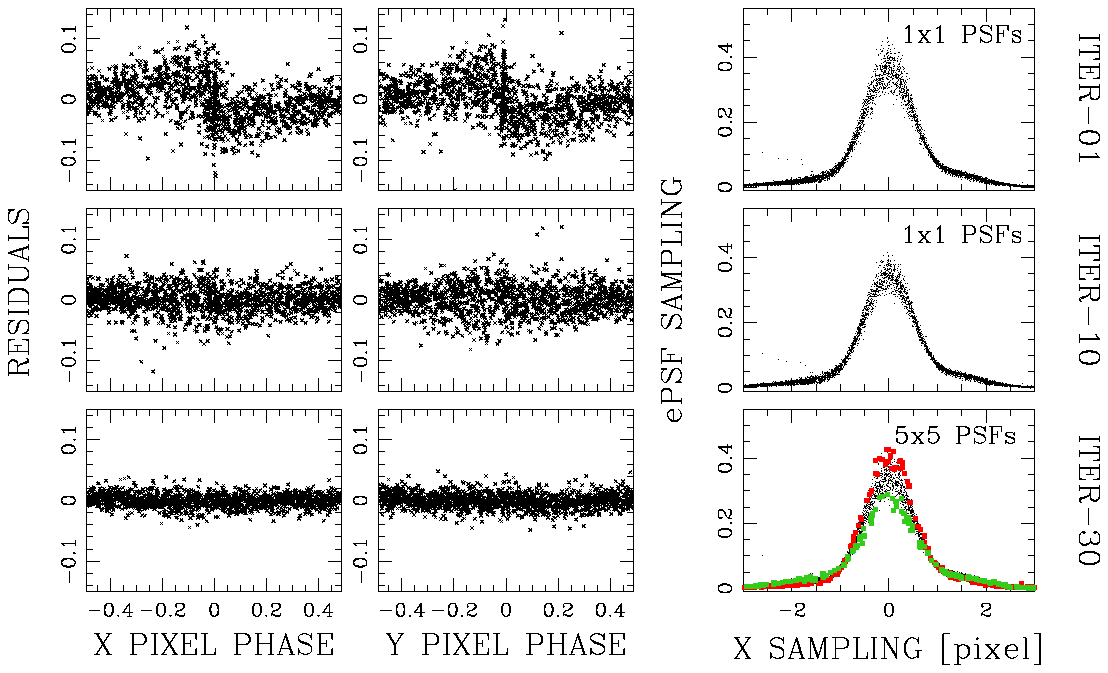}
\caption{An overview of the improvements in the ePSF models during the
  different iterations. The top-to-bottom panels display the results
  for iteration 1, 10, and 30, respectively. The left and middle
  columns show the distributions of the pixel phase errors,
  representing the position residuals of the stars relative to their
  mean positions as a function of the X/Y pixel phase. The right
  panels show the ePSF sampling at various distances from the ePSF X
  centre within a slice of $-0.25 < \delta Y < 0.25$.  In iteration 1
  and 10, a single ePSF model was employed for the entire detector; in
  iteration 30, a grid of $5 \times 5$ ePSFs was generated to account
  for the PSF variations across the detector. The green and red points
  represent the ePSF sampling for the lower-left and upper-right
  detector regions, respectively, demonstrating how the PSF peak
  changes across the detector. The figure showcases the case of the
  most undersampled filter, F070W, and focuses on modelling the ePSFs
  of detector A1. \label{fig:2}}
\end{figure*}

\subsubsection{Effective PSFs}

For each filter and detector, we derived a $5 \times 5$ grid of
$4\times$ over-sampled library ePSFs by following the procedure
developed by \citet[AK00]{2000PASP..112.1360A} for undersampled PSFs,
and employed in many other works (see, e.g,
\citealt{2004acs..rept...15A, 2006acs..rept....1A,
  2015wfc..rept....8A, 2016wfc..rept...12A,
  2016MNRAS.456.1137L,2016MNRAS.463.1780L,2016MNRAS.463.1831N,
  2023arXiv230300009L,2023MNRAS.521L..39N}).  Here, we provide a brief
description of the ePSF modelling; we refer the reader to
Paper~\citetalias{2022MNRAS.517..484N} or AK00, for a detailed
description of the method.

To obtain a well-sampled PSF model, we need to break the degeneracy
between positions and fluxes of the sources which occurs when dealing
with images whose PSFs are undersampled. The degeneracy can be broken
by constraining the positions and fluxes of a set of stars using an
iterative procedure. Initially, we extracted the first-guess positions
and fluxes of bright, unsaturated, and isolated stars in each image
using the ePSF grids obtained in
Paper~\citetalias{2022MNRAS.517..484N}\footnote{For each filter, the
\lq{closest\rq} PSF set in terms of wavelength was utilised.}. By
employing the geometric distortion (GD) solution derived in
Paper~\citetalias{2023AN....34430006G}, we determined the
six-parameter transformations between the images by cross-matching the
stars in common to different images. These transformations were
employed to create a catalogue containing the mean positions and
fluxes (transformed to a common reference system) of stars that were
measured at least three times.  This catalogue served as a master
catalogue for ePSF modelling: indeed, stars in this catalogue will be
adopted to break the degeneracy position-flux typical of undersampled
images and derive the ePSF model.

The following four steps were followed to derive the ePSF model: (1)
using the inverse GD solution from
Paper~\citetalias{2023AN....34430006G}, we transformed the positions
of the stars from the master catalogue to the reference system of each
individual image; (2) we converted each pixel value within a radius of
25 pixels from each star's centre into an estimate of the ePSF model,
projecting the individual point samplings from the original image
scale to a grid super-sampled by a factor 4 ($201 \times 201$ points);
(3) in iteration 1, we calculated the ePSF model as the
3$\sigma$-clipped average of the point sampling within a square by
$0.25 \times 0.25$ pixels$^2$ in ePSFs ($x,y$) coordinates. Starting
from iteration 2, we first subtracted from each sampling the
corresponding value of the last ePSF model, and then we calculated the
3$\sigma$-clipped average of the residuals in each $0.25 \times 0.25$
pixels$^2$ grid point. Mean residuals are then added to the last
available ePSF model and the result was smoothed with a combination of
linear, quadratic and quartic kernels; (4) using the last available
ePSF model, we re-measured the positions and fluxes of the sources in
the master list in each image and we performed the transformations
described above to obtain an updated master list to use in step (1).

We repeated steps (1)-(4) ten times assuming a single ePSF model for
the entire detector; from iteration 11 we took into account the
spatial variation of the ePSFs by dividing the image in sub-regions
and calculating the ePSF models using the sources in each region. We
gradually increased the number of subregions from $2 \times 2$ (size
of each subregion: $1024 \times 1024$ pixel$^2$) to $5 \times 5$ (size
of each subregion: $409 \times 409$ pixel$^2$).

Figure~\ref{fig:2} illustrates the improvement of the ePSF models for
the detector A1 and filter F070W (the most undersampled one) from the
initial iteration to the final iteration (30). The enhancement is
particularly evident in the distributions of the pixel phase errors,
where the distributions flatten out as the PSF model improves (panels
in the left and middle columns).  Right column panels show the
improvement of the ePSF samplings from iteration 1 to 10 when
utilising a single PSF model for the entire detector, and in the final
iteration where 25 different ePSFs are modelled. The ePSF samplings in
the lower-left and upper-right regions, represented by green and red
points respectively, demonstrate the significant variation of the PSF
peak across the detector (see Appendix~\ref{app:2}).

Middle and bottom rows of Fig.~\ref{fig:1} show the ePSFs in each
filter, demonstrating how the ePSF model changes from a filter to
another.

We make publicly available these ePSFs, which are improved with
respect to our early derivation in
Paper\,\citetalias{2022MNRAS.517..484N}, as at that time an
appropriate geometric distortion correction was not publicly
available. More details are reported in Appendix~\ref{app:1}.

\begin{figure*}
\includegraphics[height=0.4\textheight]{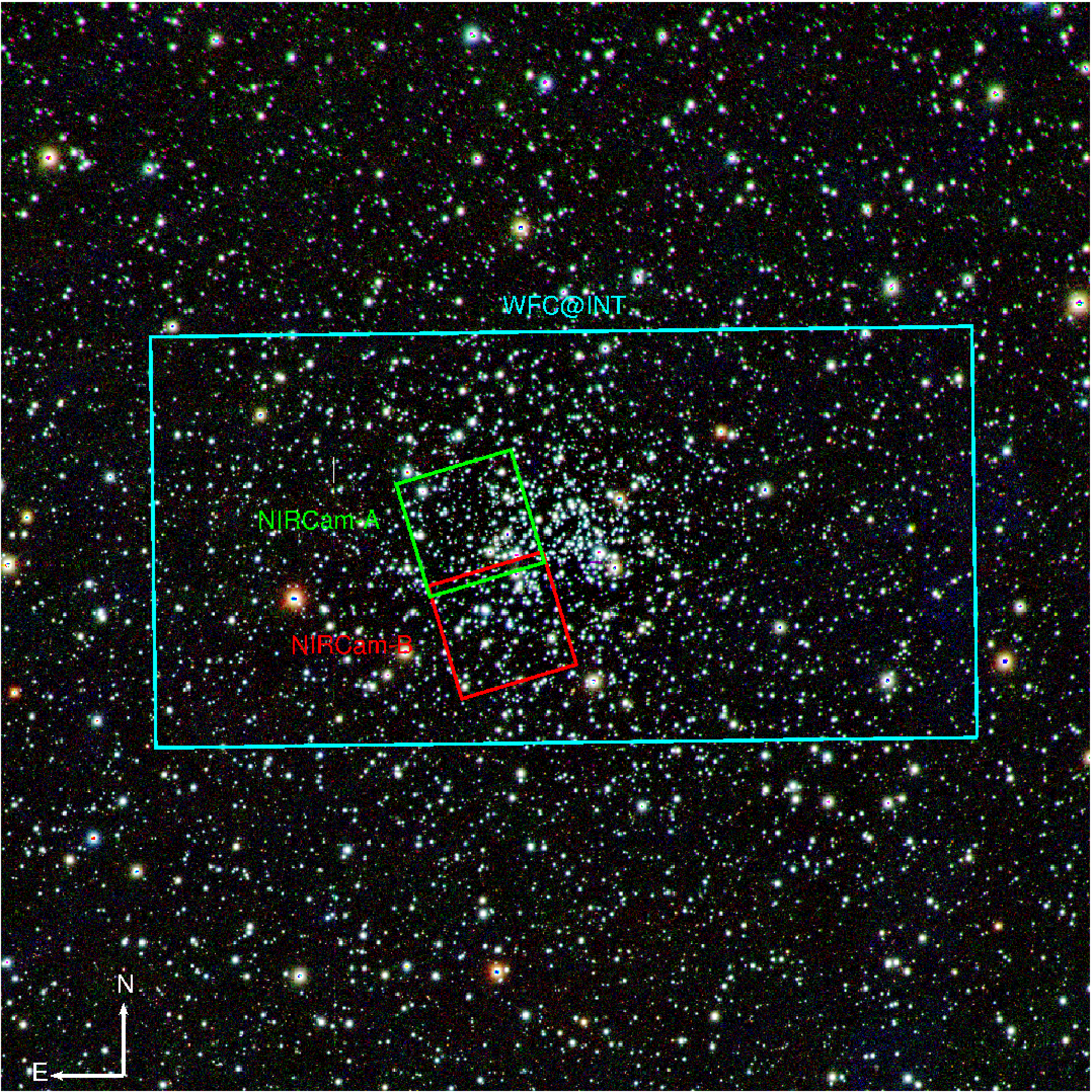}
\hspace{0.35cm}
\includegraphics[height=0.4\textheight]{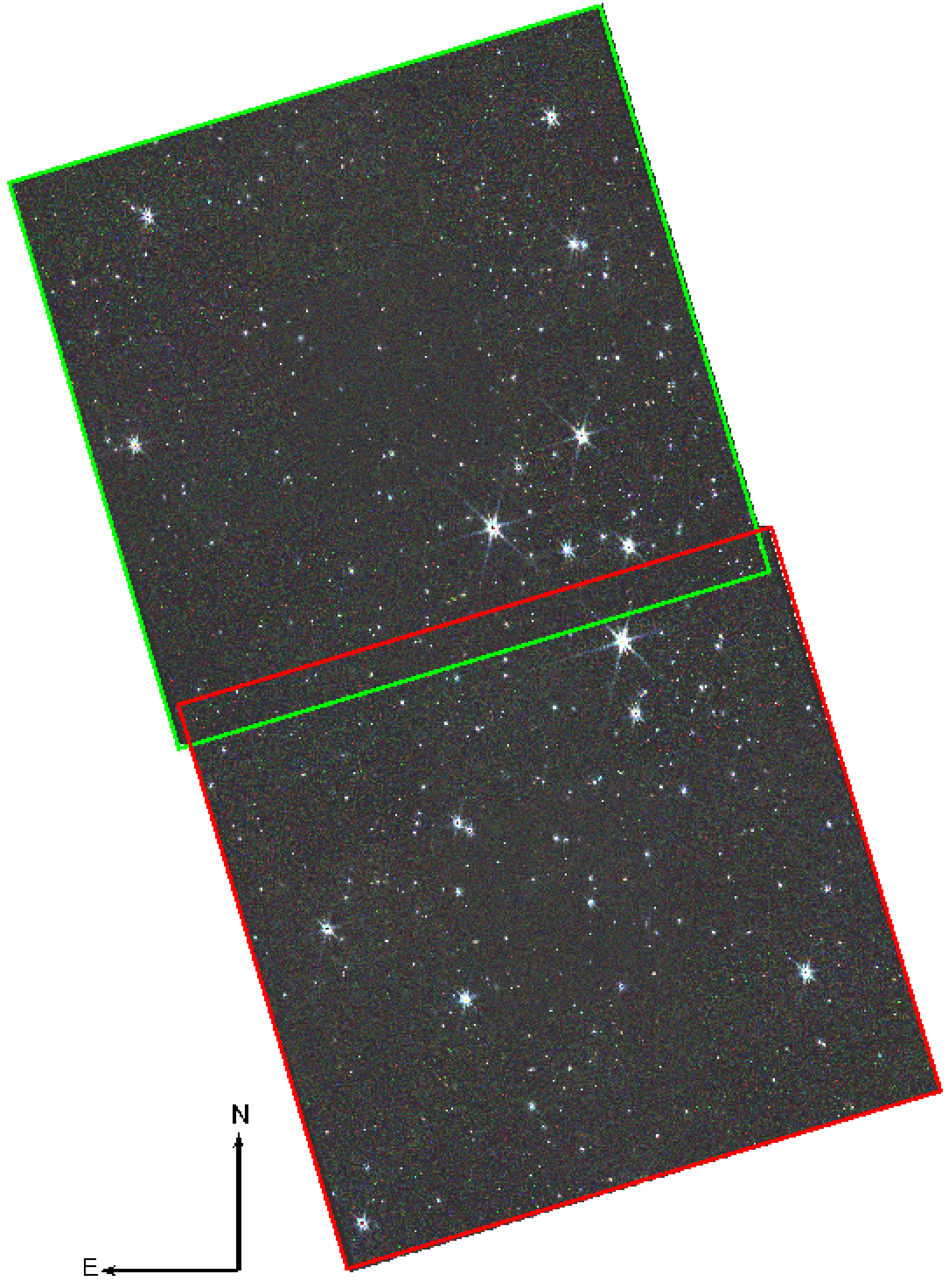}
\caption{The left panel shows a three-colour image of a $30 \times 30$
  arcmin$^2$ region centred on NGC\,2506, created by stacking $g-, r-,
  i-$sloan band images from the PS1 survey. The field of view of the
  observations obtained with NIRCam Module A and B is indicated by
  green and red squares, respectively. Additionally, the cyan
  rectangle represents the field of view of the WFC@INT detector 4
  observations employed in this study (see Sect.\,\ref{sec:wfc}). The
  right panel shows the three-colour image (F277W $+$ F356W $+$ F444W)
  generated by stacking the images obtained from the {\it JWST}
  observations employed in this work. \label{fig:3}}
\end{figure*}

%%%%%

\subsection{NGC\,2506 {\it JWST} observations}

Our target open cluster NGC\,2506 was observed by {\it JWST} with
NIRCam on November 1st, 2022 over the course of two hours, as part of
the Calibration Program CAL-1538 (PI: Gordon). The purpose of this
programme was to obtain observations of G dwarf stars for the absolute
flux calibration of {\it JWST}.  The observations consisted of a $2
\times 2$ mosaic centred in ($\alpha,\delta$)=(120.03775,$-$10.78695),
in which each dither is observed two times with \texttt{RAPID} readout
mode (2 integrations comprising 2 groups of 1 frame, $t_{\rm
  exp}=42.947~s$). Data were collected in the same ten filters (8 wide
$+$ 2 medium) employed in Sect.~\ref{sec:lmc} and shown in
Fig.~\ref{fig:1}.

Figure~\ref{fig:3} illustrates the field of view covered by the NIRCam
observations: in the left panel, a three-colour image obtained with
PanSTARRS DR1 (PS1) images
(\citealt{2016arXiv161205560C,2020ApJS..251....3M,2020ApJS..251....4W})
is shown; green and red squares represent the Modules A and B field of
views, respectively, which are also displayed in the right panel. Due
to the limited overlap between the two modules and the scarcity of
bright stars required for accurate six-parameter transformations, the
data associated with the two modules were reduced separately. The
final catalogues were combined at the end of the data reduction
process.

\begin{figure*}
\includegraphics[bb=21 224 569 694, width=0.8\textwidth]{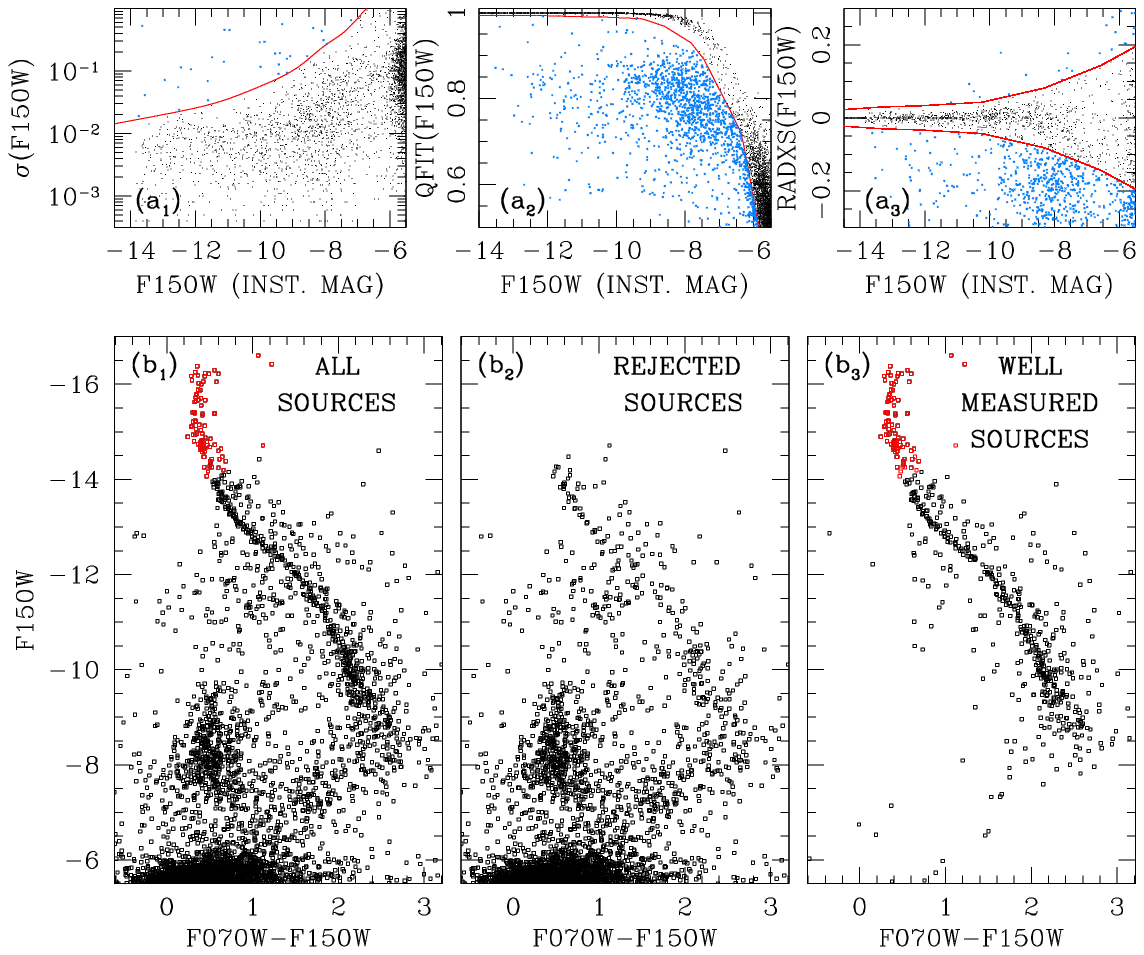}
\caption{Overview of the procedure to identify and reject bad
  measurements in the catalogues produced by the \texttt{KS2} routine.
  Panels (a$_1$), (a$_2$), and (a$_3$) show the distributions of
  photometric RMS ($\sigma$), quality-of-fit (QFIT), and RADXS for the
  filter F150W, respectively: red lines represent the threshold for
  good measurements in each parameter distribution, while the azure
  points are the rejected sources. Panels (b$_1$), (b$_2$), and
  (b$_3$) show the F150W versus (F070W$-$F150W) instrumental CMDs of all
  measured sources, the sources rejected in at least one filter, and
  the sources that passed the selection criteria in both filters,
  respectively. Saturated stars are plotted in red. The figure shows
  the sources in the catalogues obtained from Module A
  images. \label{fig:4}}
\end{figure*}

\subsubsection{Catalogues}
\label{sec:catalog}
Observations of NGC\,2506 were obtained approximately four months
after the data used to derive the library ePSFs. To account for the
time variation of the PSFs, we employed the brightest and most
isolated stars in each image to perturb the corresponding library PSFs
(for a detailed description of the procedure, refer to
\citealt{2006acs..rept....1A, 2008AJ....135.2114A,
  2018MNRAS.481.3382N}). To perturb the PSFs, we first measured the
positions and fluxes of the stars selected for the perturbation with
the library ePSF; we modelled and subtracted these stars from the
image, and calculated the average of the residuals at different
distances from the PSF centre (oversampled by a factor of 4). The mean
residuals were utilised to adjust the PSF model. We iterated this
procedure 5 times, using the last perturbed PSF each time.

We used the perturbed ePSFs to extract positions and fluxes of the
stars in each image using the NIRCam version of the \texttt{img2xym}
software, developed by \citet{2006A&A...454.1029A} for the Wide Field
Imager (WFI) @ESO/MPG 2.2m telescope. We will refer to this photometry
as \textit{first-pass} photometry. For each filter, we transformed
positions and magnitudes to a common reference frame using
six-parameter linear transformations and mean photometric zero points.
Subsequently, we used the perturbed ePSFs, the images, and the
transformations to perform the \textit{second-pass} photometry with
the \texttt{KS2} software\footnote{In this work, we utilised a
modified private copy of the \texttt{KS2} software specifically
adapted to NIRCam images.}, developed by J.~Anderson (which is a
second generation of the software \texttt{kitchen\_sync} presented in
\citealt{2008AJ....135.2114A}).

The \texttt{KS2} software enables the measurement of positions and
fluxes of bright and faint stars with high accuracy through the
simultaneous analysis of all the images (see, e.g.,
\citealt{2016ApJS..222...11S,2017ApJ...842....6B,2018MNRAS.481.3382N,2021MNRAS.505.3549S}),
employing ad-hoc masks for bright and saturated stars (recovered
during the \textit{first-pass} photometry using the frame-0 of each
image), and four different iterations in which progressively fainter
stars are measured and subtracted from the images. During the first
two iterations, we searched for sources in the F277W filter, in the
third iteration we utilised the F356W filter, and in the fourth
iteration we employed the F070W filter to identify faint \textit{blue}
sources that may not be detectable in the red filters.

Since only 45~\% of the field of view of each module is covered by
$\geq 3$ images per filter, while the remaining $\sim 55$~\% is
covered by two images (or even just one!)  and since our goal is to
identify as many weak sources as possible, the final catalogue
generated by the \texttt{KS2} software contains numerous sources that
may be identified as noise peaks in individual images. Therefore, it
is necessary to perform a "cleaning" of the catalogue. We adopted the
same selection criteria as described in
\citet{2018MNRAS.481.3382N,2019MNRAS.485.3076N}, which are based on
the following parameters: (i) the photometric error (RMS); (ii) the
quality-of-fit (QFIT), which assesses the PSF fitting of the source;
(iii) the RADXS parameter defined as in \citet{2008ApJ...678.1279B},
which enables the distinction between stellar sources, galactic-shaped
sources, and cosmic-rays/noise peaks. A detailed description of these
parameters is reported in \citet{2018MNRAS.481.3382N}, while an
example of the selection process for the F150W filter is illustrated
in panels\,(a) of Fig.~\ref{fig:4}.  Additionally, sources measured
only once and heavily contaminated by bright neighbouring stars were
excluded from the selections. The lower panels of Fig.~\ref{fig:4}
display the instrumental F150W versus F070W$-$F150W colour-magnitude
diagrams (CMDs) for all the detected sources (panel (b$_1$)), the
sources rejected by the selections (panel (b$_2$)), and the sources
that passed the selections (b$_3$)). Saturated stars (recovered from
the \textit{first-pass} photometry), that were not subjected to any
selection, are highlighted in red.

\subsubsection{Photometric calibration}

We calibrated the catalogues in each filter and module in the VEGAmag
system by using the most updated\footnote{Version 5, November 2022}
{\it JWST} photometric zero-points reported in the {\it JWST} User
documentation\footnote{https://jwst-docs.stsci.edu/jwst-near-infrared-camera/nircam-performance/nircam-absolute-flux-calibration-and-zeropoints}.
To calibrate the instrumental magnitudes, we first performed aperture
photometry (with a radius $r_{\rm ap} = 0.20$~arcsec) of the most
isolated bright stars from each \texttt{\_cal} image.
We transformed the finite-aperture fluxes ${\rm FLX}(r_{\rm
  ap}=0.20^{\prime \prime}){\rm [MJy/sr]}$ into infinite-aperture
fluxes using the Encircled Energy (EE) distributions reported in the
{\it JWST} User
documentation\footnote{https://jwst-docs.stsci.edu/jwst-near-infrared-camera/nircam-performance/nircam-point-spread-functions}
as follows:
\begin{equation}
{\rm FLX}(r_{\rm ap}= \infty){\rm [MJy/sr]} = {\rm FLX}(r_{\rm ap}=0.20^{\prime\prime}){\rm [MJy/sr]}/{\rm EE(0.20^{\prime\prime})} 
\end{equation} 
We calculated the aperture photometry calibrated magnitudes by using the
equation:
\begin{equation}
  m_{\rm ap,cal} = -2.5 \times \log{{\rm FLX[DN/s]}} + {\rm ZP}_{\rm VEGA}
\end{equation}
where the flux ${\rm FLX[DN/s]}$ is computed as
\begin{equation}
  {\rm FLX[DN/s]} = {\rm FLX}(r_{\rm ap}= \infty){\rm [MJy/sr]} / {\tt PHOTMJSR}
\end{equation}
and ${\rm ZP}_{\rm VEGA}$ and ${\tt PHOTMJSR}$ are the photometric
zero-points and the conversion factor MJy/sr to DN/s, respectively,
tabulated in the {\it JWST} User
documentation\footnote{https://jwst-docs.stsci.edu/files/182256933/182256934/1/
\ 1669487685625/NRC\_ZPs\_0995pmap.txt}.

For each filter and module, we cross-matched the aperture photometry
calibrated catalogues with the catalogue obtained in the previous
section, and, for the $N_{\rm stars}$ stars in common, we calculated
the difference
\begin{equation}
\delta m^i =  m^i_{\rm ap,cal} - m^i_{\rm inst,KS2}\,\,\,\,\,\,{\rm with\,}i=1,...,N_{\rm stars}
\end{equation}
where $m_{\rm inst,KS2}$ is the magnitude output of the \texttt{KS2}
routine. We then computed the average value of $\delta m$, which
represents the photometric zero-point to be added to the instrumental
magnitudes.

\begin{figure*}
\includegraphics[bb=17 422 497 707, width=0.9\textwidth]{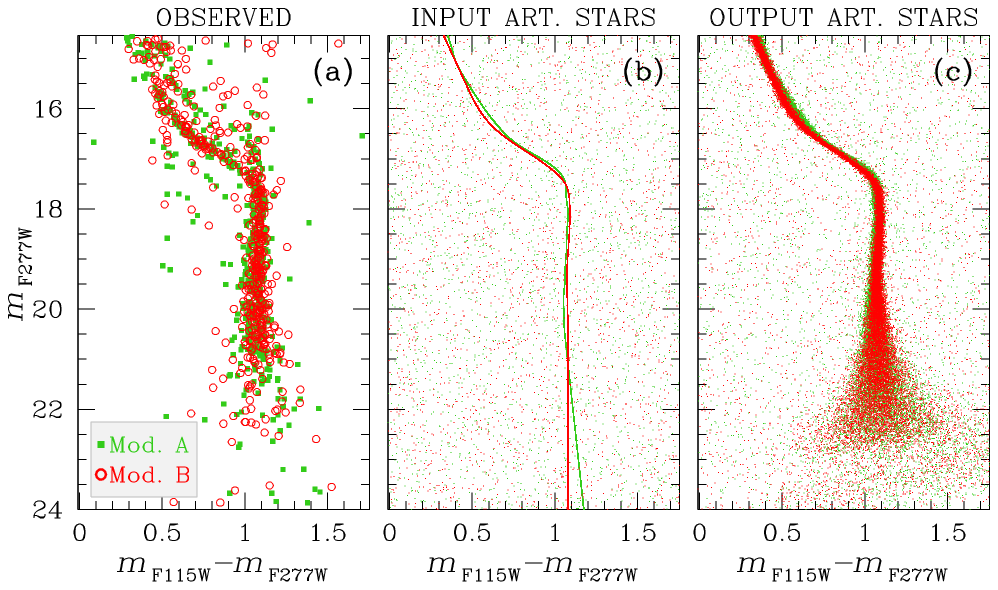}
\caption{Observed and artificial $m_{\rm F277W}$ versus $m_{\rm
    F115W}-m_{\rm F277W}$ CMDs for Module A (in green) and B (in
  red). Panel (a) shows the CMD obtained with real observations; panel
  (b) and (c) display the input and output artificial star
  CMDs. \label{fig:5}}
\end{figure*}

Due to the errors in the photometric zero-points, which are $\sim
0.02$ mag, we observed slight discrepancies between the CMDs obtained
from the stars measured in Module A and Module B images. To align the
two modules to a common reference system, we employed the following
approach: we used a small sample of stars that were detected in both
modules (consisting of 10 to 25 stars) to determine the mean
difference between the magnitudes obtained from the images of Module A
and Module B ($\delta m$(A-B)). Subsequently, we applied a correction
by subtracting (or adding) $\delta m$(A-B)/2 to the magnitudes of the
stars in Module A (or Module B). An example of the achieved results
can be seen in panel (a) of Fig.~\ref{fig:5}.

\subsubsection{Artificial stars}
\label{sec:ass}
We employed artificial stars to assess the completeness of our
catalogue across different filters (Sect.\ref{sec:completeness} and
\citealt{2018MNRAS.477.2004N} for detailed information). For each
module, we generated 50\,000 artificial stars (ASTs) within the F277W
magnitude range from $14.5$ (near the saturation limit) to $26.5$. We
created 40\,000 ASTs with a flat luminosity function in F277W and with
colours that lie along the cluster sequence in the different $m_{\rm
  F277W}$ versus $m_{\rm X}-m_{\rm F277W}$ CMDs, where X is one of the
ten available filters. Additionally, we generated other 10\,000 ASTs
with a flat luminosity function in F277W and with random colours in
the different filters, to simulate the field stars. The spatial
distribution of the ASTs was uniform across the field of view.  The
CMD of the input catalogue of ASTs is shown in panel (b) of
Fig.~\ref{fig:5}.  We used the \texttt{KS2} software to add one AST at
a time to each image with the appropriate position and flux, adopting
the same procedure used for real stars to search and measure the added
AST. The software provided output parameters for the ASTs identical to
those of real stars. Panel (c) of Fig.~\ref{fig:5} showcases the CMD
of the ASTs output of \texttt{KS2} routine.

\begin{figure}
\includegraphics[bb=37 243 358 694, width=0.4995\textwidth]{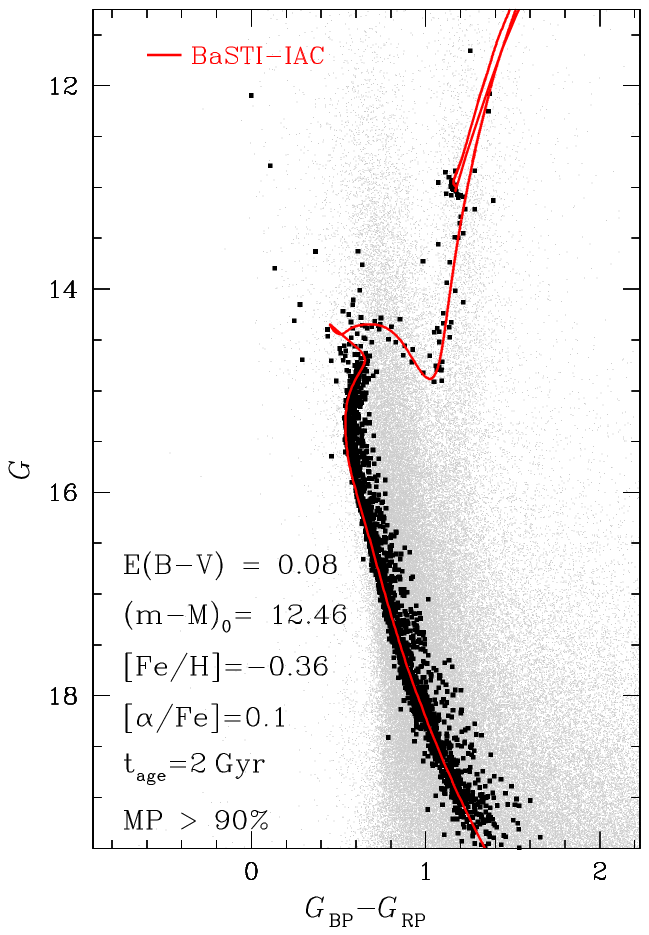}
\caption{ The $G$ versus $G_{\rm BP}-G_{\rm RP}$ CMD obtained with the
  Gaia~DR3 catalogue of the stars within 1.5 deg. from the cluster
  centre. Black and grey stars are the sources identified as {\sl
    bona-fide} cluster members and Galactic stars, respectively, based
  on the MP (threshold 90~\%); for comparison, a BaSTI-IAC 2~Gyr
  isochrone is plotted in red by adopting the parameters reported in
  the panel.\label{fig:6b}}
\end{figure}

\begin{table}
  \centering
  \caption{Cluster parameters of NGC\,2506.}
  \resizebox{0.495\textwidth}{!}{\begin{tabular}{l c c}
\hline
\hline
{\bf Parameter}     &  {\bf Value} & {\bf Reference} \\
\hline
$\alpha$(ICRS, J2015.5)~[deg.]           &   120.010  & (1)\\
$\delta$(ICRS, J2015.5)~[deg.]           & $-$10.773  & (1)\\
$l$(J2015.5)~[deg.]                      & 230.56942  & (1)\\
$b$(J2015.5)~[deg.]                      & +09.93816  & (1)\\
$d$~[pc]                                 & $3100 \pm 175$   & this work \\
$(m-M)_0$                                & $12.46 \pm 0.12 $ & this work \\
$E(B-V)$                                 & $0.08 \pm 0.01 $ & this work \\
\text{[Fe/H]}~[dex]                             & $-0.36 \pm 0.10 $  & (2) \\
\text{[$\alpha$/Fe]}~[dex]                      & $0.10 \pm 0.10 $  & (2) \\
\text{[M/H]}~[dex]                             & $-0.29 \pm 0.10 $  & (2) \\
t$_{\rm age}$~[Gyr]                        & $2.01 \pm 0.10$   & (2) \\
$r_{\rm c}$~[arcmin]                       & $2.60 \pm 0.05$   & this work \\
$r_{\rm c}$~[pc]                           & $2.34 \pm 0.05$   & this work \\
$r_{\rm t}$~[arcmin]                      & $33.0 \pm 4.3$   & this work \\
$r_{\rm t}$~[pc]                          & $29.8 \pm 3.9$   & this work \\
\hline
\multicolumn{3}{l}{{\it Notes:} \text{(1)~\citet{2021A&A...647A..19T}}; \text{(2)~\citet{2020MNRAS.499.1312K}} } \\
\end{tabular}
}
  \label{tab:1}
\end{table}

\subsection{NGC\,2506 from the Gaia archive}
\label{sec:gaia}
We adopted the Gaia~DR3 catalogue
(\citealt{2021A&A...649A...1G,2022arXiv220800211G}) to obtain
information about stars located within a radius of 1.5 degrees from
the cluster centre ($\alpha_c,\delta_c$)=($120.010,-10.773$)
(\citealt{2021A&A...647A..19T}). We calculated the membership
probabilities (MPs) using the method outlined in
\cite{2022MNRAS.511.4702G}, which considers the spatial distribution,
proper motion, and parallax of the sources. The {\sl bona-fide}
cluster members, with a membership probability $>90~\%$, are denoted
in black in Fig.~\ref{fig:6b}.  We performed a cross-match between the
{\sl bona-fide} cluster members and the catalogue by
\citet{2021AJ....161..147B} and determined the mean distance of the
cluster to be $d = 3100 \pm 175 $ pc, consistent with the value
reported by \citet{2020MNRAS.499.1312K}.  Using the Infrared Dust Maps
\footnote{https://irsa.ipac.caltech.edu/applications/DUST/}(\citealt{1998ApJ...500..525S,2011ApJ...737..103S}),
we found a mean reddening $E(B-V) = 0.08$ for the
cluster. \citet{2020MNRAS.499.1312K} determined the age, metallicity,
and $\alpha$-enhancement of NGC\,2506 through the analysis of detached
eclipsing binaries, yielding $t_{\rm age} \sim 2$~Gyr, [Fe/H] $\sim
-0.36$, and [$\alpha$/Fe] $\sim 0.10$, respectively. 
Adopting these cluster parameters (reported in Table~\ref{tab:1}), 
we plotted a 2 Gyr BaSTI-IAC isochrone (\citealt[in
  red]{2018ApJ...856..125H,2021ApJ...908..102P}) 
in the $G$ versus $G_{\rm
  BP}-G_{\rm RP}$ CMD.  
This $\alpha$-enhanced IAC-BaSTI isochrone has been derived as
follows. We have first downloaded from the BaSTI-IAC website
solar-scaled and $\alpha$-enhanced ([$\alpha$/Fe]=0.4) isochrones
(including overshooting and atomic diffusion) for [Fe/H]=$-$0.36, and
then interpolated linearly in [$\alpha$/Fe] (between 0 and 0.4) to
obtain isochrones with [Fe/H]=$-$0.36 and [$\alpha$/Fe]=0.1 (see
Table~\ref{tab:1}) -- corresponding to a total metallicity
[M/H]=$-$0.29.

\begin{figure}
\includegraphics[bb=31 308 266 691, width=0.4\textwidth]{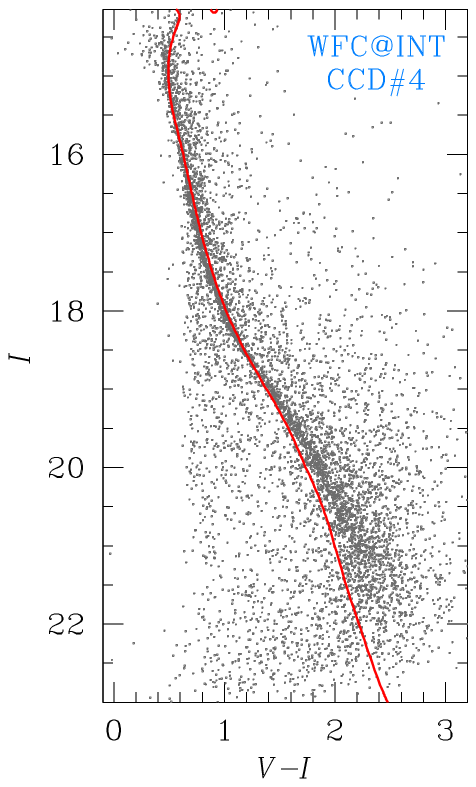}
\caption{The $I$ versus $V-I$ CMD obtained with the WFC@INT data. A
  BaSTI-IAC 2~Gyr isochrone is plotted in red.
  \label{fig:6}}
\end{figure}

\subsection{Ground-based observations of NGC\,2506}
\label{sec:wfc}

We have also taken advantage of data obtained with the Wide Field
Camera (WFC) mounted at the prime focus of the 2.5m Isaac Newton
Telescope (INT) for Programme I9/2003B (PI: Rosenberg). The
observations were conducted between January 21 and 22, 2004. The
dataset includes the following exposures: $3 \times 300$~s $+$ $2
\times 600$~s $V$-Harris images and $1 \times 30$~s $+$ $2 \times
900$~s $i$-Sloan images. All observations were taken at an airmass
between 1.32 and 1.48 and the typical seeing was around 1-1.2
arcsec. For this work, only images associated with CCD\#4, which
contains the {\it JWST} observations, were used (as shown in
Fig.~\ref{fig:3}).

We reduced the dataset by using empirical PSFs and the software
developed by \citet{2006A&A...454.1029A} adapted for WFC@INT
images. We corrected the geometric distortion by following the
procedure adopted by \citet{2022MNRAS.515.1841G}. We calibrated the
$V, i$ magnitudes into $V$-Johnson and $I$-Cousins respectively, by
using the homogeneous photometry published by
P.~.B.~Stetson\footnote{https://www.cadc-ccda.hia-iha.nrc-cnrc.gc.ca/en/community/-STETSON/homogeneous/}
(\citealt{2000PASP..112..925S}).

The resulting $I$ versus $V-I$ CMD from WFC@INT data is shown in
Fig.~\ref{fig:6}, together with a 2~Gyr BaSTI-IAC isochrone adopting
the same chemical composition, reddening, and distance as in the
comparison with the Gaia data.  The isochrone matches the pattern of
the CMD distribution of the observed stars reasonably well for $I
\lesssim 19.5$.

\section{Astrometric and photometric analyses}
\label{sec:astrophoto}

\subsection{Proper motions and membership probabilities}

To extend the membership study to magnitudes beyond the reach of Gaia
(i.e., for stars fainter than $G\simeq19$), we used proper motions
derived from {\it JWST} and INT positions. We discriminated between
cluster members and field stars, following the methodology employed in
previous studies by our research group (see, e.g.,
\citealt{2003AJ....126..247B, 2006A&A...454.1029A,
  2010A&A...513A..50B, 2014A&A...563A..80L, 2015A&A...573A..70N,
  2016MNRAS.455.2337N}).  To achieve this, we calculated the
displacements of stars between two different epochs after transforming
their positions to a common reference system using 6-parameter global
transformations. The first epoch was represented by the WFC@INT data,
with a mean epoch of $t_{\rm I}=2004.06$, while for the second epoch
we took advantage of the NIRCam catalogues, with $t_{\rm
  II}=2022.84$. We determined the relative displacements of the stars,
referring them to the average motion of the cluster, over a time span
$\Delta t_{\rm II-I} = 18.78$ years.

We computed the membership probability (MP) for each of these faint
stars in the catalogues, following an approach similar to that
described by \cite{2022MNRAS.511.4702G}. Only proper motions were
considered in the calculation of the MPs, without incorporating
spatial and parallax terms. Specifically, the spatial term was
neglected due to the limited field of view of {\it JWST} observations,
which allowed us to assume a uniform distribution of the cluster's
stars within the observed area, while parallaxes are not available for
these faint stars.

\begin{figure}
\includegraphics[bb=31 242 382 714, width=0.4995\textwidth]{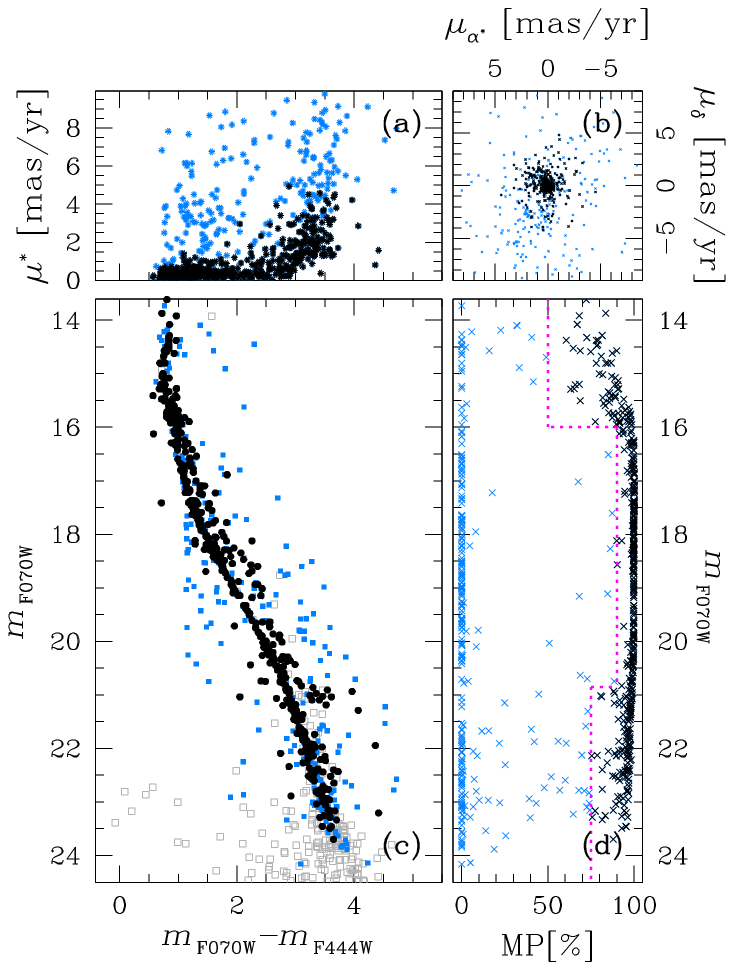}
\caption{Cluster membership analysis using the WFC@INT and NIRCam@{\it
    JWST} data. Panel (a) shows the distribution of proper motions
  relative to the cluster's mean motion, as a function of the $m_{\rm
    F070W}-m_{\rm F444W}$ colour; panel (b) presents the relative
  proper motions of the stars common to both datasets. Panel (c) is
  the $m_{\rm F070W}$ versus $m_{\rm F070W}-m_{\rm F444W}$ CMD of the
  stars in the fields of the two modules: grey squares are the stars
  not measured in the WFC@INT data set. Panel (d) showcases the MP
  distribution; the magenta line denotes the MP threshold used in this
  work. In all panels, cluster members are represented by black
  points, while field stars are denoted by azure
  points. \label{fig:8}}
\end{figure}

Figure~\ref{fig:8} illustrates the selection of cluster members based
on MPs. In the range $16.0 < m_{\rm F070W} \leq 20.5$, stars with
MP$>90~\%$ were chosen as cluster members. For saturated stars and
faint stars with $m_{\rm F070W} > 20.5$, we relaxed the selection
criteria and identified cluster members with MP$>50~\%$ and
MP$>75~\%$, respectively, as indicated in panel (d) (magenta line).
Relative proper motions are shown in panels (a) (as a function of the
colour $m_{\rm F070W}-m_{\rm F444W}$) and (b), while panel (c) shows
the $m_{\rm F070W}$ versus $m_{\rm F070W}-m_{\rm F444W}$ CMD for
cluster members (black) and field objects, mainly Galactic field stars
(azure).

\subsection{Completeness}
\label{sec:completeness}

By using the ASTs obtained as described in Section~\ref{sec:ass}, we
determined the completeness in different regions of the field of view
and for different magnitudes.  The completeness is calculated as the
ratio between the number of recovered stars and the number of injected
stars, $N_{\rm rec}/N_{\rm inj}$.  We considered an AST as recovered
if the differences in position satisfy $|x_{\rm in}-x_{\rm out}|<1$~px
and $|y_{\rm in}-y_{\rm out}|<1$~px; additionally, we applied the
following magnitude conditions: $|m_{\rm F277W,in}-m_{\rm
  F277W,out}|<0.5$ for stars measured in iteration 1 or 2, $|m_{\rm
  F356W,in}-m_{\rm F356W,out}|<0.5$ for ASTs measured in iteration 3,
and $|m_{\rm F070W,in}-m_{\rm F070W,out}|<0.5$ for sources measured in
iteration 4. These conditions reflect the procedure used to identify
real stars. To account for the rejections described in
Sect.~\ref{sec:catalog}, we applied to the ASTs the same selections
used for real stars, and included them in the computation of $N_{\rm
  rec}$.

\begin{figure*}
\includegraphics[bb=17 421 587 702, width=0.9\textwidth]{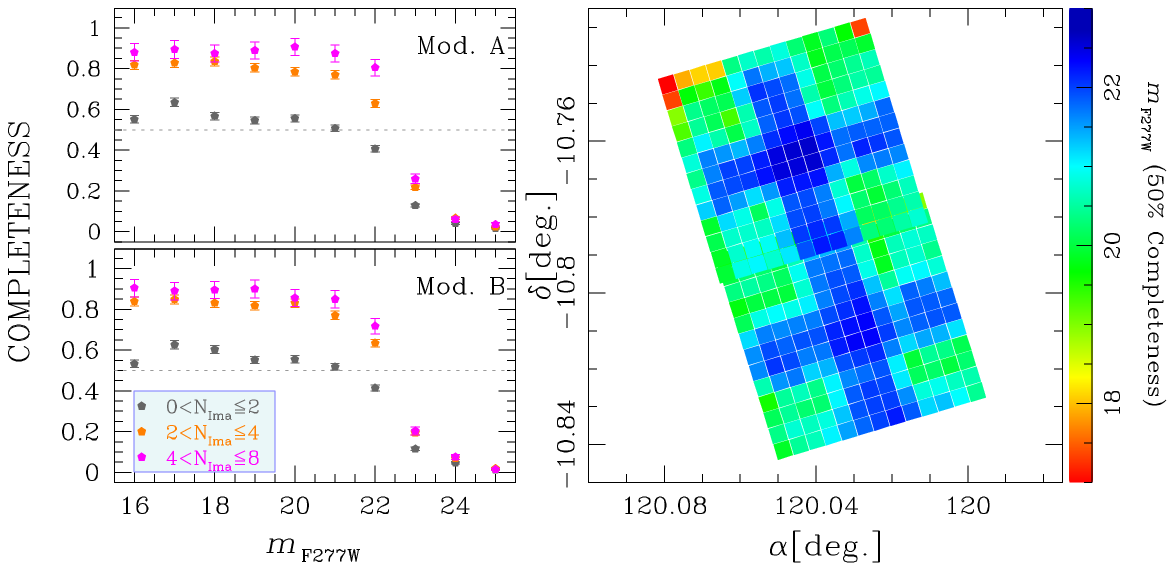}
\caption{Analysis of the completeness of our catalogues: the left
  panels show the completeness as a function of the $m_{\rm F277W}$
  magnitude for Module A (upper panel) and B (bottom panel). Magenta,
  orange, and dark grey points represent the completeness distribution
  in the case the stars appear in less than 3, between 2 and 4, and
  between 4 and 8 images, respectively.  The right panel shows the
  completeness as a function of the position: we calculated the
  $m_{\rm F277W}$ limit magnitude for which the completeness reaches
  50~\% in squares of 15.5$\times$15.5 arcsec$^2$, coloured following
  the colour-bar on the right.
\label{fig:7}}
\end{figure*}

Figure~\ref{fig:7} shows the completeness as a function of the
magnitude and position. The left panels demonstrate how completeness
varies with magnitude for different image coverages: when stars are
observed just once or two times, completeness is lower than 60~\%, and
drops below 50~\% at $m_{\rm F277W} \sim 21$ (dark grey points). For
stars measured in at least 3 images the completeness is 20-30~\%
higher, and it is lower than 50~\% at $m_{\rm F277W} \sim 22.5$
(magenta and orange points). The right panel shows the dependence of
completeness on position due to varying image coverage: the blue/azure
squares of each module are the regions covered by $\geq 4$ images, and
50~\% completeness is reached at magnitude $m_{\rm F277W} \sim
21.5$--$22.5$; green/red regions are areas covered by $\leq 2$ images,
where completeness reaches 50~\% at $m_{\rm F277W} \sim 20.0$--$21.0$.

\subsection{Colour-magnitude diagrams}
\begin{figure*}
\includegraphics[bb=17 517 583 708, width=0.95\textwidth]{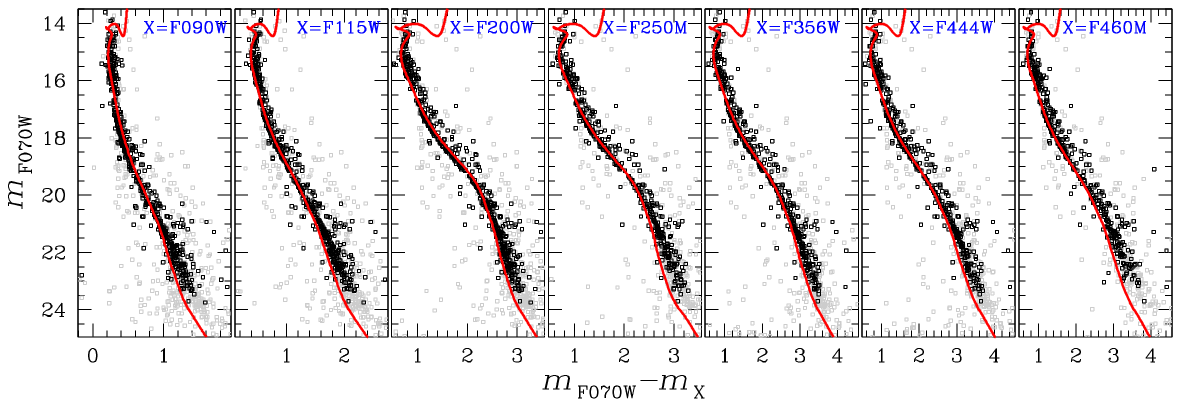} \\
\includegraphics[bb=17 517 583 708, width=0.95\textwidth]{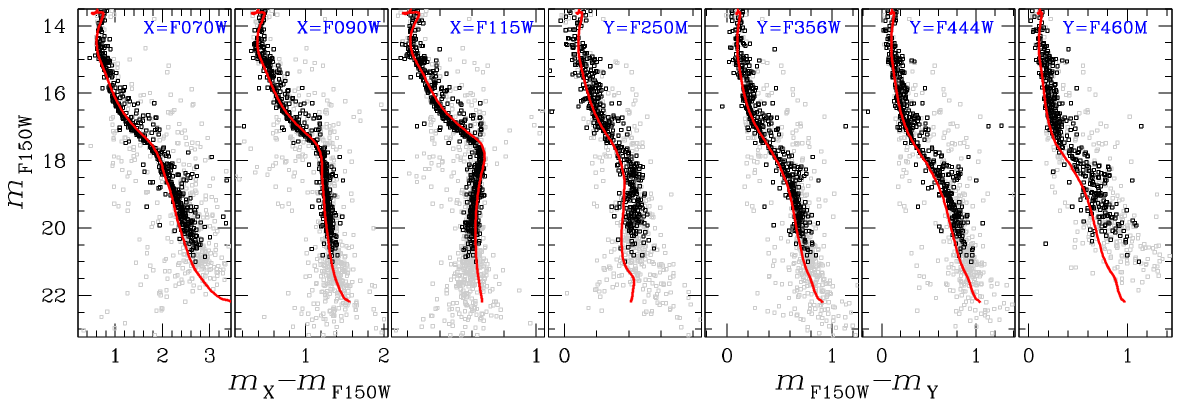} \\
\includegraphics[bb=17 517 583 708, width=0.95\textwidth]{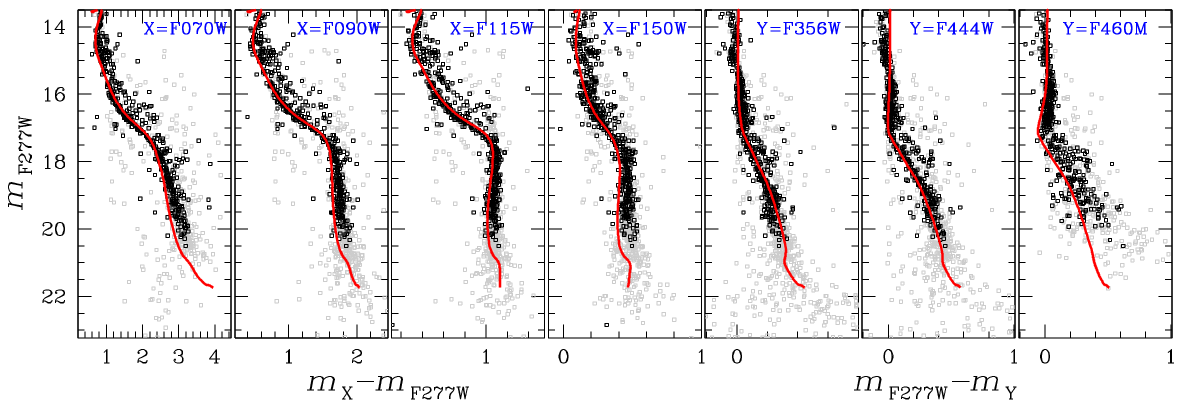}
\caption{Some of the possible CMDs obtained with the available {\it
    JWST} photometry in different filters: panels in top row show the
  $m_{\rm F070W}$ vs $(m_{\rm F070W}-m_{\rm X})$ CMDs with X=F090W,
  F115W, F200W, F250M, F356W, F444W, F460M; panels in the middle row
  are the $m_{\rm F150W}$ vs $(m_{\rm X}-m_{\rm F150W})$ CMDs with
  X=F070W, F090W, F115W, and the $m_{\rm F150W}$ vs $(m_{\rm
    F150W}-m_{\rm Y})$ CMDs with Y=F250M, F356W, F444W, F460M; in the
  bottom row are shown the the $m_{\rm F277W}$ vs $(m_{\rm X}-m_{\rm
    F277W})$ CMDs with X=F070W, F090W, F115W, F150W and the $m_{\rm
    F277W}$ vs $(m_{\rm F277W}-m_{\rm Y})$ CMDs with Y=F356W, F444W,
  F460M.  2~Gyr BaSTI-IAC isochrones are plotted in red.
\label{fig:9}}
\end{figure*}

Figure~\ref{fig:9} shows a rundown of the CMDs with the ten available
filters: grey and black points are the stars that passed the quality
and membership selections, respectively.  Naturally, these CMDs have
different levels of completeness along the MS.  As already done in
Sect.~\ref{sec:gaia} and \ref{sec:wfc}, we superimposed 2~Gyr
BaSTI-IAC isochrones (in red) to the {\it JWST} CMDs. The lower mass
limit of the isochrones is 0.1~$M_{\odot}$.  Some filters
combinations, as for example $m_{\rm F150W}$ vs $(m_{\rm F115W}-m_{\rm
  F150W})$ or $m_{\rm F150W}$ vs $(m_{\rm F150W}-m_{\rm F356W})$,
allow us to reach a magnitude limit $m_{\rm F150W} \sim 22$ with
SNR$\sim 10$; according to the models, this magnitude limit
corresponds to masses below $0.1M_{\odot}$ on the main sequence.

\begin{figure*}
\includegraphics[bb=21 364 561 696, width=0.95\textwidth]{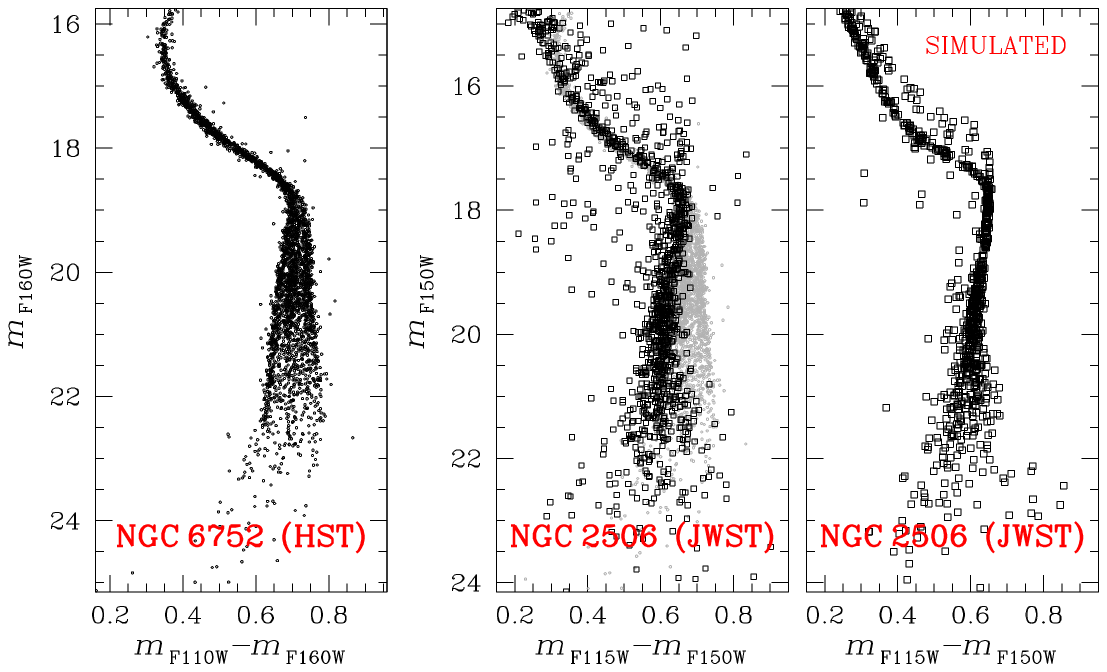} \\
\caption{Comparison between lower MS of NGC\,6752 (three stellar
  populations) and NGC\,2506 (single population) is similar IR
  filters. Left panel shows the {\it HST} $m_{\rm F160W})$ vs $(m_{\rm
    F110W}-m_{\rm F160W})$ CMD of NGC\,6752, for which three different
  low-MSs are well detectable. Middle panel is a comparison between
  the CMD of NGC\,6752 (grey points) and the $m_{\rm F150W})$ vs
  $(m_{\rm F115W}-m_{\rm F150W})$ CMD of NGC\,2506 (black points),
  that demonstrates that the lower-MS of the open cluster agrees with
  the presence of a single population. This is also demonstrated by
  the simulated CMD shown in the right panel.
\label{fig:9d}}
\end{figure*}

 We investigated the lower MS of NGC~2506 to check the presence of
 broadening due to chemical variations. The analysis is shown in
 Fig.~\ref{fig:9d}: we compared the $m_{\rm F150W}$ vs $(m_{\rm
   F115W}-m_{\rm F150W})$ CMD of NGC\,2506 against the {\it HST}
 $m_{\rm F160W}$ vs $(m_{\rm F110W}-m_{\rm F160W})$ CMD of the old,
 metal-poor, massive globular cluster NGC\,6752 (leftmost panel,
 Scalco et al. in prep.). As reported by \citet{2019MNRAS.484.4046M},
 the multiple lower MSs in NGC\,6752 are due to the variations of
 Oxygen between the stars belonging to different stellar
 populations. We shifted the CMD of NGC\,6752 in colour and magnitude
 in such a way that the "knee" of the MS overlaps the one of the
 NGC\,2506's CMD: the middle panel of Fig.~\ref{fig:9d} demonstrates
 that the lower MS of NGC\,2506 agrees with a single sequence of
 NGC\,6752 and that no chemical variation is expected among the
 low-mass stars of this cluster. As a further test, we simulated the
 CMD of a single population by selecting randomly 600 ASTs (about the
 number of observed stars in the MS) from the sample of ASTs; we also
 simulated 57.5~\% MS-MS binaries (with a flat mass-ratio
 distribution, see Sect.~\ref{sec:binmf}). The simulated CMD is
 reported in right-hand panel of Fig.~\ref{fig:9d}: it demonstrates
 that the observed CMD agrees with the simulated CMD of a single
 stellar population.

\begin{figure}
\includegraphics[bb=17 160 284 711, width=0.495\textwidth]{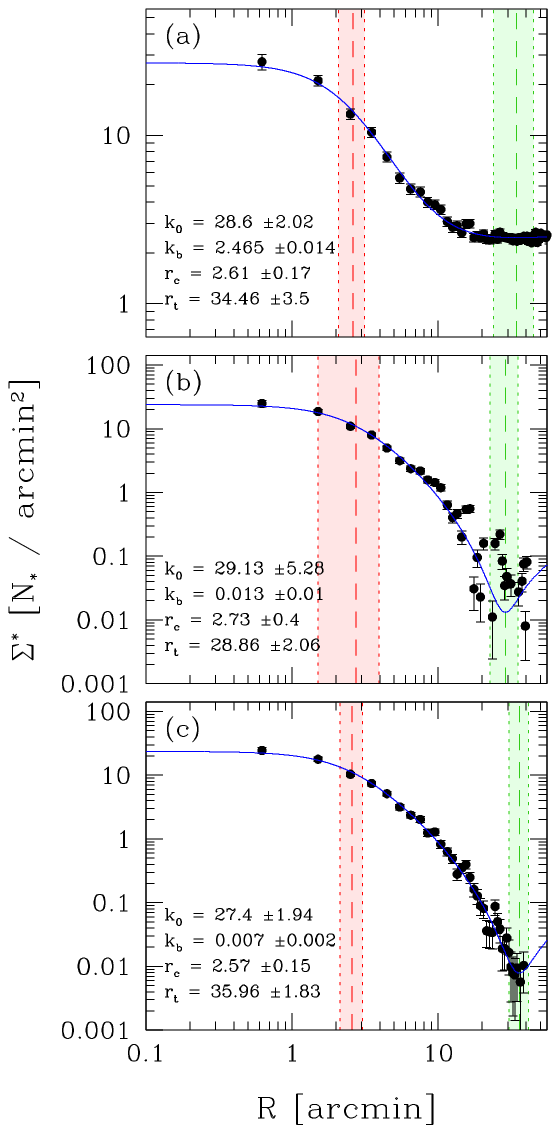}
\caption{Radial stellar density profiles of NGC\,2506, calculated for
  stars with magnitudes between $G=14$ and $G=18$ in the Gaia~DR3
  catalogue. The profiles are obtained by dividing the cluster region
  into annuli of 1 arcmin width. The three panels display different
  approaches: panel (a) shows the density profile modelling both
  cluster and Galactic field stars; in panel (b), the mean density of
  background stars is subtracted from the density profile of panel
  (a); panel (c) showcases the density profile with each star's
  contribution weighted by MP. The observed density profile is
  represented by black points, and a King profile function is fitted
  to the three profiles (blue lines). The vertical dashed lines in red
  and green denote the core and tidal radii, respectively, with shaded
  regions representing the corresponding $\pm 1\sigma$ errors. The
  parameters of the King profile function are listed within each
  panel. \label{fig:11}}
\end{figure}

\section{Radial density profile}
\label{sec:radprof}
We computed the radial stellar density profile of NGC\,2506 by using
the Gaia~DR3 catalogue. The results are presented in
Fig.~\ref{fig:11}. To obtain the density profile, we divided the area
within a radius of 70~arcmin into a series of annuli with a 1~arcmin
width.  For each annulus, we computed the number of stars per
arcmin$^2$ ($\Sigma^\star$) and the mean radial distance of the stars
from the centre ($R$). We restricted our analysis to stars with
magnitudes between $G=14$ and $G=18$, which --in the average Galactic
field-- should have a completeness close to 100\,\%
(\citealt{2020MNRAS.497.4246B, 2022MNRAS.509.6205E}).

To fit the density profile distributions shown in Fig.~\ref{fig:11},
we employed a King profile (\citealt{1962AJ.....67..471K}), described
by the function:
\begin{equation}
\Sigma^\star[R] = k_{\rm b} + k_{\rm 0} \left( \frac{1}{\sqrt{1+R^2/r_{\rm c}^2}} - \frac{1}{\sqrt{1+r_{\rm t}^2/r_{\rm c}^2}} \right)^2 
\end{equation}
where $k_{\rm b}$ and $k_{\rm 0}$ represent the background and central
stellar densities, respectively, $r_{\rm c}$ is the core radius, and
$r_{\rm t}$ denotes to the cluster tidal radius.

We analysed the density profile of NGC\,2506 in three different cases,
shown in the three panels in Fig.~\ref{fig:11}. In case (a), we fitted
a density profile that model the contribution of both the cluster
stars and the Galactic field stars. In case (b), we subtracted the
mean density of background stars from the calculated stellar density
in each annulus. The background density was determined in an annulus
ranging from $R=70$ to $R=90$~arcmin from the cluster centre. In case
(c), we fitted a density profile in which the contribution of each
star is weighted by the MP.  Excluding the background density $k_{\rm
  b}$, which is higher in case (a) due to the inclusion of Galactic
field stars in the fit, the other parameters, such as $k_0$, $r_{\rm
  c}$, and $r_{\rm t}$, show consistent values, and they agree within
3$\sigma$. The weighted mean values we obtained are as follows: $k_0 =
28.1 \pm 0.9$ stars/arcmin$^2$, $r_{\rm c} = 2.60 \pm 0.05$ arcmin,
and $r_{\rm t} = 33.0 \pm 4.3$ arcmin.

The core radius we measured is smaller than the value reported by
\citet{2019MNRAS.490.1383R} based on WFI and Gaia~DR2 data, although
they agree within 3.2$\sigma$. However, our core radius agrees with
the value found by \citet{2013MNRAS.432.1672L} for stars with $V<17$
magnitude. The central density and tidal radius obtained in our study
are higher than those reported by \citet[16.58 stars/arcmin$^2$ and 12
  arcmin, respectively]{2019MNRAS.490.1383R}, with a difference $>10
\sigma$. This discrepancy may be attributed to differences in the
selection of the cluster members and/or the different density profile
models used in the analysis. Additionally, \citet{2013MNRAS.432.1672L}
found a smaller tidal radius ($\sim 19$~arcmin). These discrepancies
can also be explained by the presence of tidal tails extending out to
1 degree from the cluster centre, as reported by \citet[even if we did
  not detect any well defined tail as in this
  work]{2020ApJ...894...48G}.

\section{Binaries and mass functions}
\label{sec:binmf}
By combining Gaia~DR3 and {\it JWST} data, we performed an analysis to
determine the fraction of photometric binaries, as well as the
luminosity and mass functions for MS stars. This analysis considered
stars located at different distances from the cluster centre and
within various mass intervals.

\subsection{Main sequence Photometric binaries}

At the distance of NGC\,2506 essentially all binaries are unresolved.
An unresolved binary system composed of two MS stars with magnitudes
$m_1$ and $m_2$, corresponding to fluxes $F_1$ and $F_2$, can be treated as a point-like source with a total magnitude given by:
\begin{equation}
m_{\rm bin}=m_1-2.5 \log{\left(1+\frac{F_2}{F_1}\right)}
\end{equation}
The ratio $F_2/F_1$ is strongly correlated with the mass ratio $q =
M_2/M_1$, where $M_1$ represents the mass of the primary (more
massive) star and $M_2 \leq M_1$. In the case of $M_2 = M_1$, the
magnitude of the binary system is $\sim 0.75$ magnitudes brighter than
that of the primary component. Thus, the $q=1$ case serves as a limit
for the redder and brighter section of the MS, where MS-MS binaries
are located.

\begin{figure*}
\includegraphics[bb=20 369 576 689, width=0.895\textwidth]{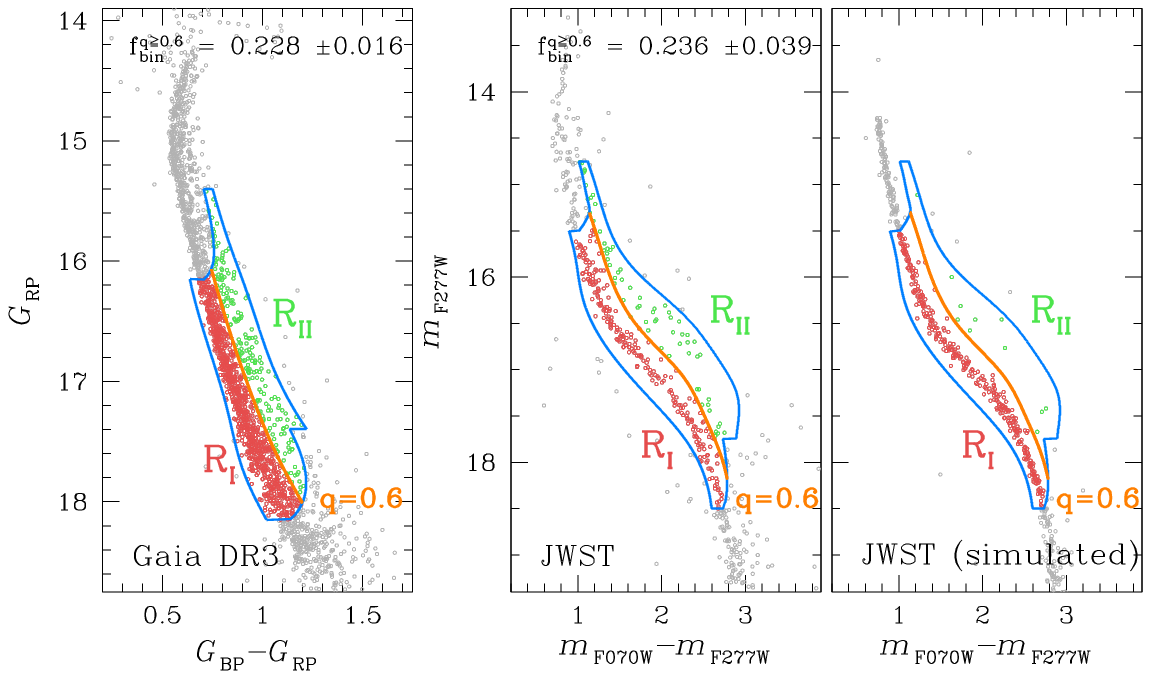}
\caption{ Procedure employed to determine the MS binary fraction from
  the Gaia~DR3 $G_{\rm RP}$ vs $(G_{\rm BP}-G_{\rm RP})$ CMD (left
  panel), and the {\it JWST} $m_{\rm F277W}$ vs. $(m_{\rm
    F070W}-m_{\rm F277W})$ CMD (middle and right panels). Each CMD is
  divided in two regions: (i) R$_{\rm I}$ contains the majority of MS
  and low mass-ratio binary stars (red points); (ii) R$_{\rm II}$
  includes the binaries with mass ratios between $q=0.6$ and $q=1.0$,
  while also accounting for the uncertainties in colours and
  magnitudes (green points). The right panel shows the simulated {\it
    JWST} CMD, which was used to assess the impact of photometric
  errors and blending on the computation of the binary fraction. The
  primary mass interval covered in the Gaia~DR3 data analysis is $0.75
  \lesssim M \lesssim 1.10~M_{\odot}$, while in the case of {\it JWST}
  data we studied binaries in the primary mass interval $0.45 \lesssim
  M \lesssim 1.05~M_{\odot}$.
  \label{fig:10a}}
\end{figure*}

To estimate the fraction of binaries, we employed the {\it JWST} and
Gaia~DR3 catalogues following the approach described by
\citet{2012A&A...540A..16M,2016MNRAS.455.3009M} and
\citet{2023A&A...672A..29C}, and shown in Fig.~\ref{fig:10a}. Briefly,
we defined a region R$_{\rm I}$ encompassing MS stars and binaries
with $q < q^{\rm lim}$, where $q^{\rm lim}$ denotes the mass ratio
below which is difficult to discriminate between single and binary
stars in the CMD. In this study, we used $q^{\rm lim} = 0.6$,
represented as an orange line in Fig.~\ref{fig:10a}, which was
determined by fitting the BaSTI-IAC isochrones discussed in the
previous sections.  To define the bluer limit of R$_{\rm I}$, we
calculated the standard deviation of the colour ($\sigma_{\rm col}$)
at various magnitudes along the MS. We considered all the stars whose
colour is $>{\rm col}_{\rm fid}-2.5\sigma_{\rm col}$, where ${\rm
  col}_{\rm fid}$ represents the colour of the MS fiducial line. In
the case of Gaia CMD, we used the $G_{\rm BP}-G_{\rm RP}$ colour,
while for the {\it JWST} data we adopted the $m_{\rm F070W}-m_{\rm
  F277W}$ wide colour baseline. The region R$_{\rm I}$ is confined
within the magnitude intervals $16.15 \leq G_{\rm RP} \leq 18.15 $ in
the case of Gaia CMD, and $15.50 \leq m_{\rm F277W} \leq 18.50 $ for
the {\it JWST} data\footnote{It is worth noting that $m_{\rm F277W} =
18.50$ was chosen as the faint end due to the rapid decline in
completeness in the F070W filter at fainter magnitudes, making it
nearly impossible to measure the binary fraction using other filter
combinations where the MS is vertical.}. All the stars in the region
R$_{\rm I}$ are plotted in red in Fig.~\ref{fig:10a}. The region
R$_{\rm II}$ is defined as the area between the lines determined by
binary systems with $q=0.6$ (represented by the orange line) and $q=1$
shifted by $2.5\sigma_{\rm col}$. The faint and bright limits of
R$_{\rm II}$ correspond to the positions of binary systems with mass
ratios $0 \leq q \leq 1$. In Fig.~\ref{fig:10a}, stars located within
the region R$_{\rm II}$ are shown in green.

In the case of Gaia~DR3, we assumed a completeness of the catalogue
equal to 100~\%, and we calculated the fraction of binaries with mass
ratios $q \geq 0.6$ using the following formula:
\begin{equation}
f_{\rm bin}^{q \geq 0.6}=\frac{\sum_i{w^i_{\rm MP}}}{\sum_j{w^j_{\rm MP}}}\,\,\,\,\,{\rm with\,\,\,}i=1,...,N_{\rm II};\,\,\,j=1,...,N_{\rm I} 
\end{equation}
where $w_{\rm MP}$ is a weight ranging from 0 to 1 corresponding to
membership probabilities from 0~\% to 100~\%, and $N_{\rm I}$ and
$N_{\rm II}$ represent the number of sources within the region R$_{\rm
  I}$ and R$_{\rm II}$, respectively.

In the case of {\it JWST} data, we took into account the completeness
of our catalogue as well as the effects of photometric errors and
blending, which can increase the number of sources in region R$_{\rm
  II}$.  To address this issue, we made use of ASTs to simulate a CMD
as follows: for each real star in our catalogue that passed the
quality selections and had membership probabilities $>90~\%$, we
randomly selected a star from the ASTs catalogue located within a
radius $r_{\rm AST}=3$ arcsec from the target star, with a maximum
difference in F277W magnitude of $\delta m_{\rm F277W, AST}=1$
mag. The simulated CMD is shown in the right panel of
Fig.~\ref{fig:10a}. For the {\it JWST} data, we estimated the fraction
of MS binaries with mass ratios $q \geq 0.6$ using the following
equation:
\begin{equation}
f_{\rm bin}^{q \geq 0.6}=\frac{\sum_i{w^i_{\rm MP}/c^i}}{\sum_j{w^j_{\rm MP}/c^j}}-\frac{N^{\rm AST}_{\rm II}}{N^{\rm AST}_{\rm I}}\,\,\,\,\,{\rm with\,\,\,}i=1,...,N_{\rm II};\,\,\,j=1,...,N_{\rm I} 
\end{equation}
where $w_{\rm MP}$ is the weight defined earlier, $c$ denotes the
completeness associated with each star, and $N^{\rm AST}_{\rm I}$ and
$N^{\rm AST}_{\rm II}$ correspond to the number of simulated stars
that fall within R$_{\rm I}$ and R$_{\rm II}$, respectively.  For the
region covered by the Gaia~DR3 catalogue ($0 \lesssim R \lesssim
35$~arcmin), we obtained $f_{\rm bin}^{q \geq 0.6}=0.228 \pm 0.016$ in
the primary mass interval $0.75 \lesssim M \lesssim 1.10~M_{\odot}$,
while for the region covered by {\it JWST} data ($0.2 \lesssim R
\lesssim 4.6$~arcmin) we measured $f_{\rm bin}^{q \geq 0.6}=0.236 \pm
0.039$ in the primary mass range $0.45 \lesssim M \lesssim
1.05~M_{\odot}$.  Supposing a flat mass-ratio distribution
(\citealt{2012A&A...540A..16M}), the total fraction of MS-MS binary
stars with q$>0$ is $\sim 0.575$.

\begin{figure}
\includegraphics[bb=20 237 338 693, width=0.495\textwidth]{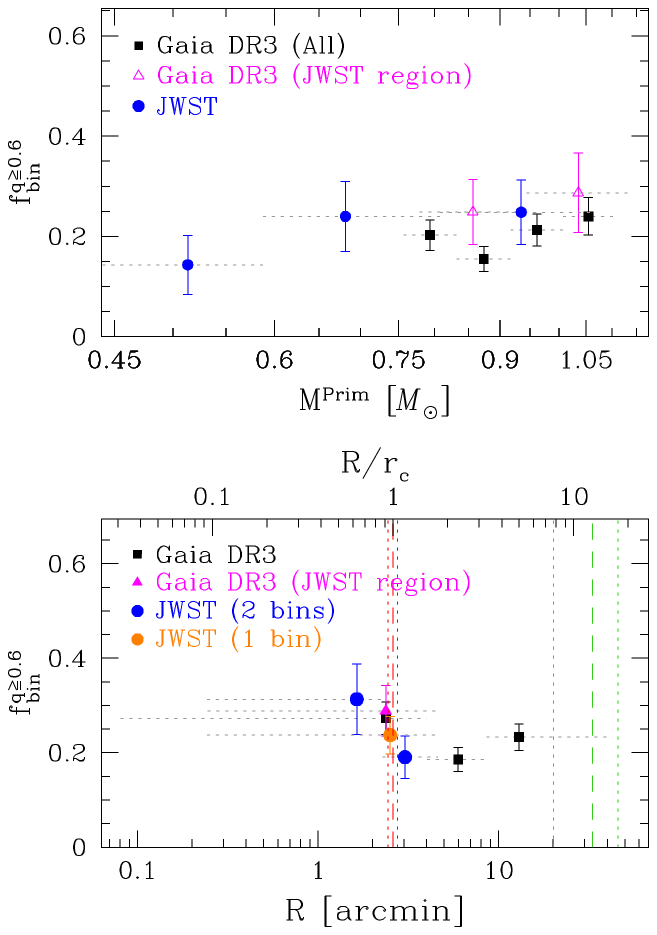}
\caption{ Analysis of the binary fraction as a function of the mass of
  the primary star (top panel) and of the radial distance from the
  cluster centre (bottom panel). Black squares, magenta triangles, and
  blue circles refer to the analysis performed with the entire
  Gaia~DR3 catalogue, the Gaia~DR3 catalogue limited to the {\it JWST}
  region, and the {\it JWST} data, respectively. The orange point in
  the bottom panel is referred to the analysis performed with {\it
    JWST} in only one radial interval.
\label{fig:10b}}
\end{figure}

We investigated the binary fraction as a function of the mass of the
primary star. To measure the binary fraction in different mass
intervals, we followed the procedure described earlier while adjusting
the lower and upper magnitude limits of the R$_{\rm I}$ region. The
results are shown in the top panel of Fig.\ref{fig:10b} and in
Table~\ref{tab:2}: the binary fraction measured in the entire field
using Gaia~DR3 data is reported in black for four different mass
intervals, the binary fraction measured with Gaia in the overlapping
region with {\it JWST} is shown in magenta, and the binary fraction
calculated using {\it JWST} data down to 0.45~$M_{\odot}$ is
represented in blue.  We performed a $\chi^2$ test to assess the
flatness of the $f_{\rm bin}^{q \geq 0.6}$ distribution as a function
of the mass of the primary star. Considering all the data points and
their corresponding errors from the top panel of Fig.~\ref{fig:10b},
we constructed a histogram with three bins and calculated the $\chi^2$
between the observed and expected frequencies, resulting in
$\chi^2=0.67$. Subsequently, using the $\chi^2$-distribution, we
determined the P-value, which represents the probability that the
distribution is flat. The obtained P-value was $\sim 72~\%$.

Furthermore, we computed the radial distribution of the binary
fraction for both the Gaia and {\it JWST} datasets, as illustrated in
the bottom panel of Fig.~\ref{fig:10b} and in Table~\ref{tab:2}. For
this distribution, we divided the Gaia and {\it JWST} samples into
radial bins, each containing an equal number of stars. The three
measurements obtained with Gaia are denoted by black squares, while
the two measurements from {\it JWST} are represented by blue
circles. These measurements exhibit a radial trend, with a higher
concentration of binaries in the central region.  We also examined the
consistency between the Gaia and {\it JWST} measurements in the
overlapping region of the two datasets, displayed by the magenta
triangle and orange circle. The P-value test yielded a probability of
$\sim 41~\%$ that the distribution is flat, suggesting a plausible
radial trend in the binary fraction.  \\ \\

\begin{table}
  \centering
  \caption{Analysis of photometric binaries in NGC\,2506.}
  \resizebox{0.495\textwidth}{!}{\begin{tabular}{l c c c}
\hline
\hline
\multicolumn{4}{c}{\text{$f^{q \geq 0.6}_{\rm bin}$} vs \text{$M^{\rm prim}$}} \\
\hline
{\bf Region} & {\bf $M^{\rm prim}$ interval} & {\bf Mean $M^{\rm prim}$} & {\bf $f^{q\geq 0.6}_{\rm bin}$} \\
             & ($M_{\odot}$)         & ($M_{\odot}$)    &                               \\
\hline
Gaia~(All)     & 0.76--0.83 & 0.79 & $0.203 \pm 0.030$ \\
               & 0.83--0.92 & 0.87 & $0.155 \pm 0.025$ \\
               & 0.92--1.01 & 0.96 & $0.213 \pm 0.032$ \\
               & 1.01--1.10 & 1.05 & $0.240 \pm 0.037$ \\
Gaia~({\it JWST} region)    & 0.78--0.94 & 0.86 & $0.249 \pm 0.065$ \\
                            & 0.94--1.13 & 1.04 & $0.287 \pm 0.079$ \\
{\it JWST}     & 0.44--0.59 & 0.51 & $0.143 \pm 0.059$ \\
               & 0.59--0.81 & 0.68 & $0.240 \pm 0.070$ \\
               & 0.81--1.06 & 0.94 & $0.248 \pm 0.059$ \\
\hline
\hline
\multicolumn{4}{c}{\text{$f^{q \geq 0.6}_{\rm bin}$} vs \text{$R$}} \\
\hline
{\bf Region} & {\bf $R$ interval} & {\bf Mean $R$} & {\bf $f^{q\geq 0.6}_{\rm bin}$} \\
             & (arcmin)         & (arcmin)    &                               \\
\hline
Gaia~(All)     & 0.08--4.00  & 2.38 & $0.273 \pm 0.035$ \\
               & 4.00--8.58  & 5.98 & $0.185 \pm 0.025$ \\
               & 8.58--40.48 & 12.97 & $0.233 \pm 0.028$ \\
Gaia~({\it JWST} region)  & 0.24--4.61 & 2.37 & $0.289 \pm 0.053$ \\
{\it JWST} (1 bin) & 0.24--4.61 & 2.51 & $0.237 \pm 0.040$ \\
{\it JWST} (2 bin) & 0.24--2.26 & 1.64 & $0.313 \pm 0.075$ \\
                   & 2.26--4.61 & 3.03 & $0.191 \pm 0.045$ \\
\hline
\end{tabular}

% Binary fraction vs M1
%
%%%% GAIA %%%%%
% 1.102224435 1.007801543 1.054640139 0.2400529681 0.03738617611
% 1.007801543 0.9173567638 0.9619201711 0.2126980455 0.03209509352
% 0.9173567638 0.8324295174 0.8739524054 0.1547416465 0.02467631112
% 0.8324295174 0.7567549211 0.7934311367 0.2025572008 0.03043875806
%%%% GAIA JWST %%%%%
% 1.131829976 0.9438400674 1.035789046 0.287002287 0.07948115271
% 0.9438400674 0.778393273 0.8571754908 0.2486763098 0.06515238932
%%%%  JWST %%%%%
% 1.05799192 0.8077731313 0.9347853078 0.2481426522 0.06420774606
% 0.8077731313 0.587568321 0.6814985727 0.2398409244 0.06952811564
% 0.587568321 0.4404528537 0.5133690745 0.1432319946 0.05898360848

% Binary fraction vs R

%GAIA  0.08044878282 3.995303601 2.382191161  0.2729621902 0.03450860171
%GAIA  4.004865547 8.582465496 5.979681809  0.185331691 0.02529986502
%GAIA  8.597342292 40.4800862 12.9662683  0.2331316012 0.0281159874
%Read lines 1 to 1 from binarie_radmodAB.dat
%GAIA JWST 0.2437621636 4.613042201 2.372892506  0.2886272741 0.05320348301
%Read lines 1 to 2 from binarie_rad.dat
%Read lines 1 to 2 from binarie_radART.dat
%JWST 0.2432385212 2.26079769 1.640091197  0.3131571285 0.07503354406
%JWST 2.271775454 4.612556099 3.030716228  0.1905474516 0.04520085261
%Read lines 1 to 1 from binarie_radmodAB.dat
%Read lines 1 to 1 from binarie_radmodABART.dat
%JWST 0.2432385212 4.612556099 2.510242717  0.237172699 0.03978408661
%
}
  \label{tab:2}
\end{table}

\begin{figure*}
\includegraphics[bb=17 178 576 689, width=0.995\textwidth]{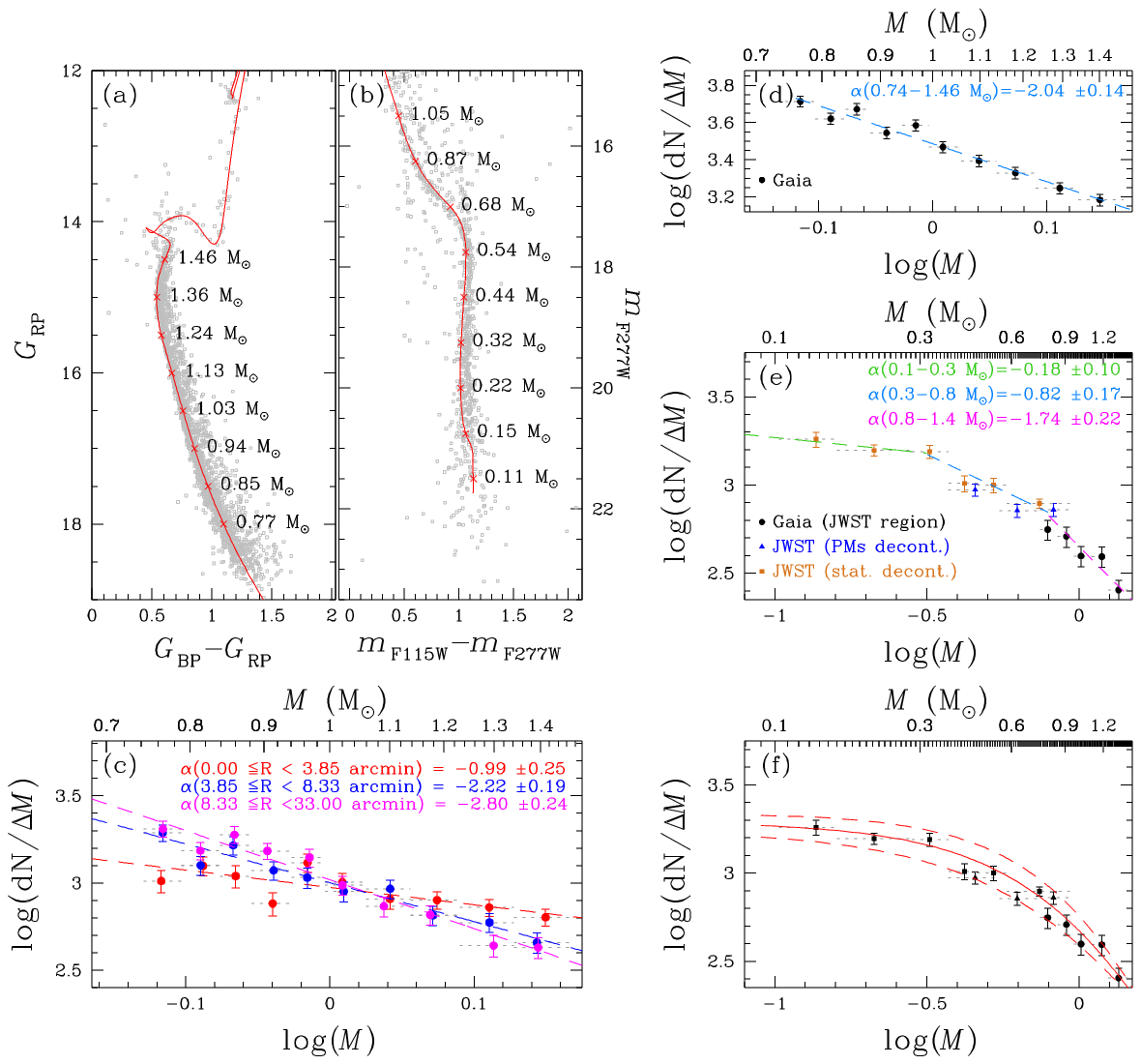}
\caption{An overview of the procedure to calculate the MS MFs from
  Gaia~DR3 and {\it JWST} data. Panels (a) and (b) show the $G_{\rm
    RP}$ vs $(G_{\rm BP}-G_{\rm RP})$ and $m_{\rm F277W}$ vs $(m_{\rm
    F115W}-m_{\rm F277W})$ CMDs, respectively, with superimposed in
  red the BaSTI-IAC isochrones marking the mass intervals on the
  MS. Panels (c) and (d) show the MFs calculated in three radial
  intervals and over the whole field, respectively, using the Gaia~DR3
  dataset; slopes ($\alpha$) of the fitted straight lines are reported
  inside the panels. Panels (e) and (f) show the MF obtained using the
  stars in the {\it JWST} region and combining Gaia and {\it JWST}
  data. For {\it JWST} data we adopted two different approaches based
  on the PMs and statistical decontamination. In panel (e) the MF is
  fitted with three different straight lines ($\alpha$ slopes are
  reported in the panel) corresponding to different mass intervals.
  In panel (f) we adopted a logistic function (see text for details).
\label{fig:12}}
\end{figure*}

\begin{table}
  \centering
  \caption{MFs of MS stars in NGC\,2506.}
  \resizebox{0.495\textwidth}{!}{\begin{tabular}{l c c c}
\hline
\hline
\multicolumn{4}{c}{Global MF from Gaia~DR3 catalogue} \\
\hline
{\bf R interval}    & {\bf $M$ interval} & {\bf Mean $M$} & {\bf $({\rm dN}/\Delta M)$} \\
(arcmin)   &    ($M_{\odot}$)    & ($M_{\odot}$)    &                               \\
\hline
0.00--33.00 &  0.74--0.79        &   0.77         &  $5168 \pm 335$ \\
           &  0.79--0.84        &   0.81         &  $4175 \pm 288$ \\
           &  0.84--0.88        &   0.86         &  $4699 \pm 331$ \\
           &  0.88--0.94        &   0.91         &  $3505 \pm 244$ \\
           &  0.94--0.99        &   0.97         &  $3846 \pm 274$ \\
           &  0.99--1.06        &   1.02         &  $2938 \pm 209$ \\
           &  1.06--1.14        &   1.10         &  $2474 \pm 178$ \\
           &  1.14--1.23        &   1.18         &  $2132 \pm 151$ \\
           &  1.23--1.34        &   1.29         &  $1765 \pm 126$ \\
           &  1.34--1.46        &   1.40         &  $1526 \pm 109$ \\
\hline
\hline
\multicolumn{4}{c}{Local MFs from Gaia~DR3 catalogue} \\
\hline
{\bf R interval}    & {\bf $M$ interval} & {\bf Mean $M$} & {\bf $({\rm dN}/\Delta M)$} \\
(arcmin)   &    ($M_{\odot}$)    & ($M_{\odot}$)    &                               \\
\hline
0.00--3.85 &  0.74--0.79 &  0.76 & $   1026 \pm    149$ \\
           &  0.79--0.84 &  0.82 & $   1254 \pm    158$ \\
           &  0.84--0.88 &  0.86 & $   1094 \pm    160$ \\
           &  0.88--0.94 &  0.91 & $    763 \pm    114$ \\
           &  0.94--0.99 &  0.97 & $   1307 \pm    160$ \\
           &  0.99--1.06 &  1.02 & $   1012 \pm    123$ \\
           &  1.06--1.14 &  1.10 & $    810 \pm    102$ \\
           &  1.14--1.23 &  1.19 & $    798 \pm     92$ \\
           &  1.23--1.34 &  1.29 & $    725 \pm     81$ \\
           &  1.34--1.46 &  1.41 & $    636 \pm     70$ \\
3.85--8.33 &  0.74--0.79 &  0.77 & $   1939 \pm    205$ \\
           &  0.79--0.84 &  0.81 & $   1263 \pm    158$ \\
           &  0.84--0.88 &  0.86 & $   1642 \pm    196$ \\
           &  0.88--0.94 &  0.91 & $   1181 \pm    142$ \\
           &  0.94--0.99 &  0.96 & $   1075 \pm    145$ \\
           &  0.99--1.06 &  1.02 & $    893 \pm    115$ \\
           &  1.06--1.14 &  1.10 & $    926 \pm    109$ \\
           &  1.14--1.23 &  1.18 & $    652 \pm     83$ \\
           &  1.23--1.34 &  1.29 & $    593 \pm     73$ \\
           &  1.34--1.46 &  1.39 & $    456 \pm     60$ \\
8.33--33.00 &  0.74--0.79 &  0.77 & $   2043 \pm    211$ \\
            &  0.79--0.84 &  0.81 & $   1532 \pm    174$ \\
            &  0.84--0.88 &  0.86 & $   1891 \pm    210$ \\
            &  0.88--0.94 &  0.91 & $   1526 \pm    161$ \\
            &  0.94--0.99 &  0.97 & $   1399 \pm    165$ \\
            &  0.99--1.06 &  1.02 & $    971 \pm    120$ \\
            &  1.06--1.14 &  1.09 & $    735 \pm     97$ \\
            &  1.14--1.23 &  1.17 & $    657 \pm     84$ \\
            &  1.23--1.34 &  1.30 & $    438 \pm     63$ \\
            &  1.34--1.46 &  1.39 & $    427 \pm     58$ \\
\hline
\hline
\multicolumn{4}{c}{Local MFs from Gaia~DR3 \& {\it JWST} catalogues ($0.24 \leq R \leq 4.61$~arcmin)} \\
\hline
{\bf Region}    & {\bf $M$ interval} & {\bf Mean $M$} & {\bf $({\rm dN}/\Delta M)$} \\
   &    ($M_{\odot}$)    & ($M_{\odot}$)    &                               \\
\hline
{\it JWST} stat.~decont. &  0.10--0.16 &  0.14 & $   1816 \pm    179$ \\
                         &  0.16--0.28 &  0.21 & $   1570 \pm    114$ \\
                         &  0.28--0.37 &  0.32 & $   1549 \pm    131$ \\
                         &  0.37--0.46 &  0.42 & $   1022 \pm    106$ \\
                         &  0.46--0.58 &  0.52 & $    999 \pm     93$ \\
                         &  0.58--0.93 &  0.74 & $    787 \pm     47$ \\
{\it JWST} PMs decont. &  0.37--0.55 &  0.46 & $    939 \pm     73$ \\
                       &  0.55--0.74 &  0.63 & $    715 \pm     61$ \\
                       &  0.74--0.93 &  0.82 & $    724 \pm     61$ \\
Gaia~DR3      &  0.74--0.85 &  0.79 & $    559 \pm     73$ \\
              &  0.85--0.96 &  0.91 & $    511 \pm     68$ \\
              &  0.96--1.10 &  1.02 & $    396 \pm     53$ \\
              &  1.10--1.25 &  1.19 & $    393 \pm     52$ \\
              &  1.25--1.46 &  1.35 & $    254 \pm     34$ \\
\hline
\end{tabular}
}
  \label{tab:4}
\end{table}

\begin{figure*}
    \begin{minipage}[t]{0.52\textwidth}
     \begin{subfigure}{\linewidth}
        \includegraphics[bb=22 282 347 677, height=0.40\textheight]{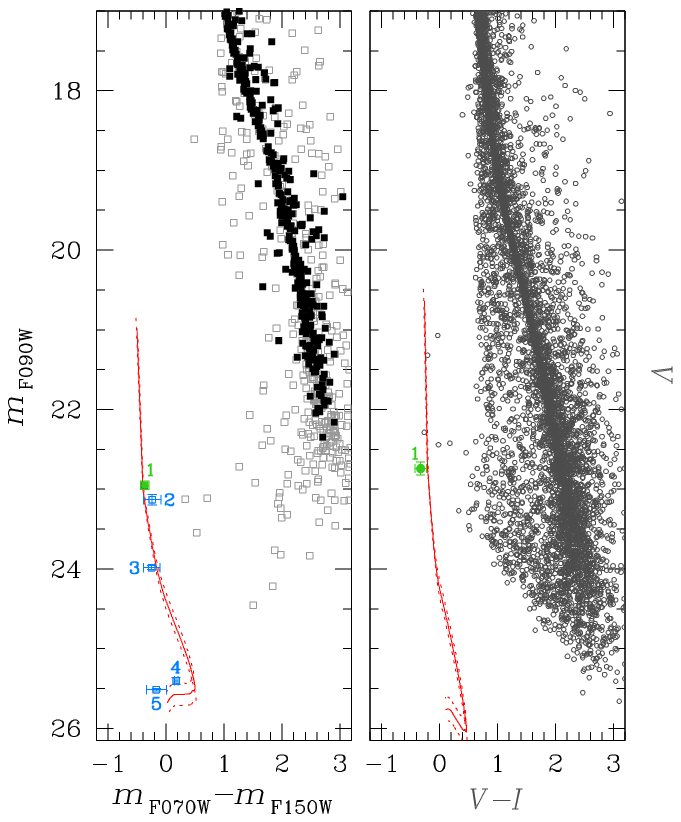}
     \end{subfigure} 
    \end{minipage}
    \hfill
  \begin{minipage}[t]{0.45\textwidth}
     \begin{subfigure}{\linewidth}
       \includegraphics[width=0.125\textheight]{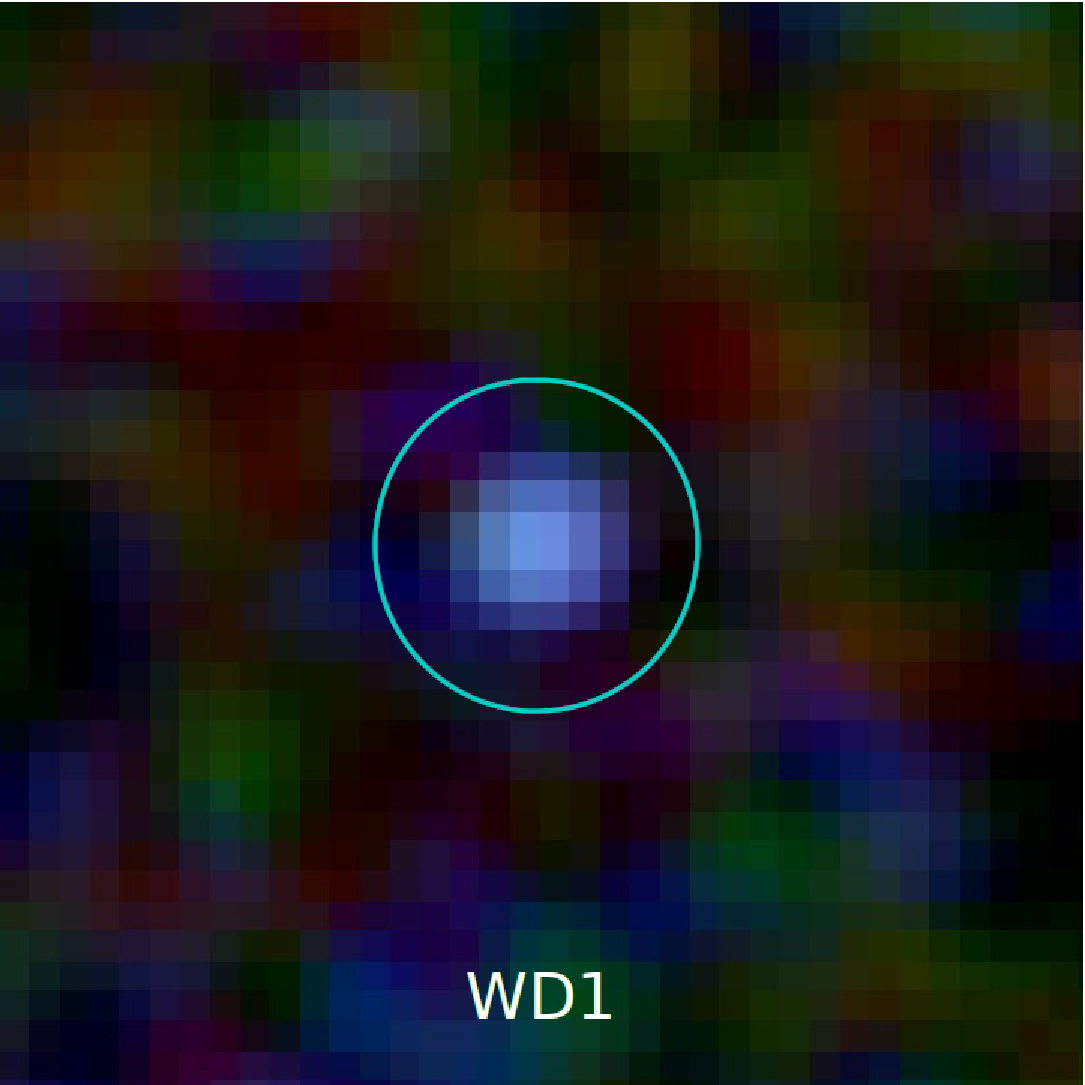}
       \includegraphics[width=0.125\textheight]{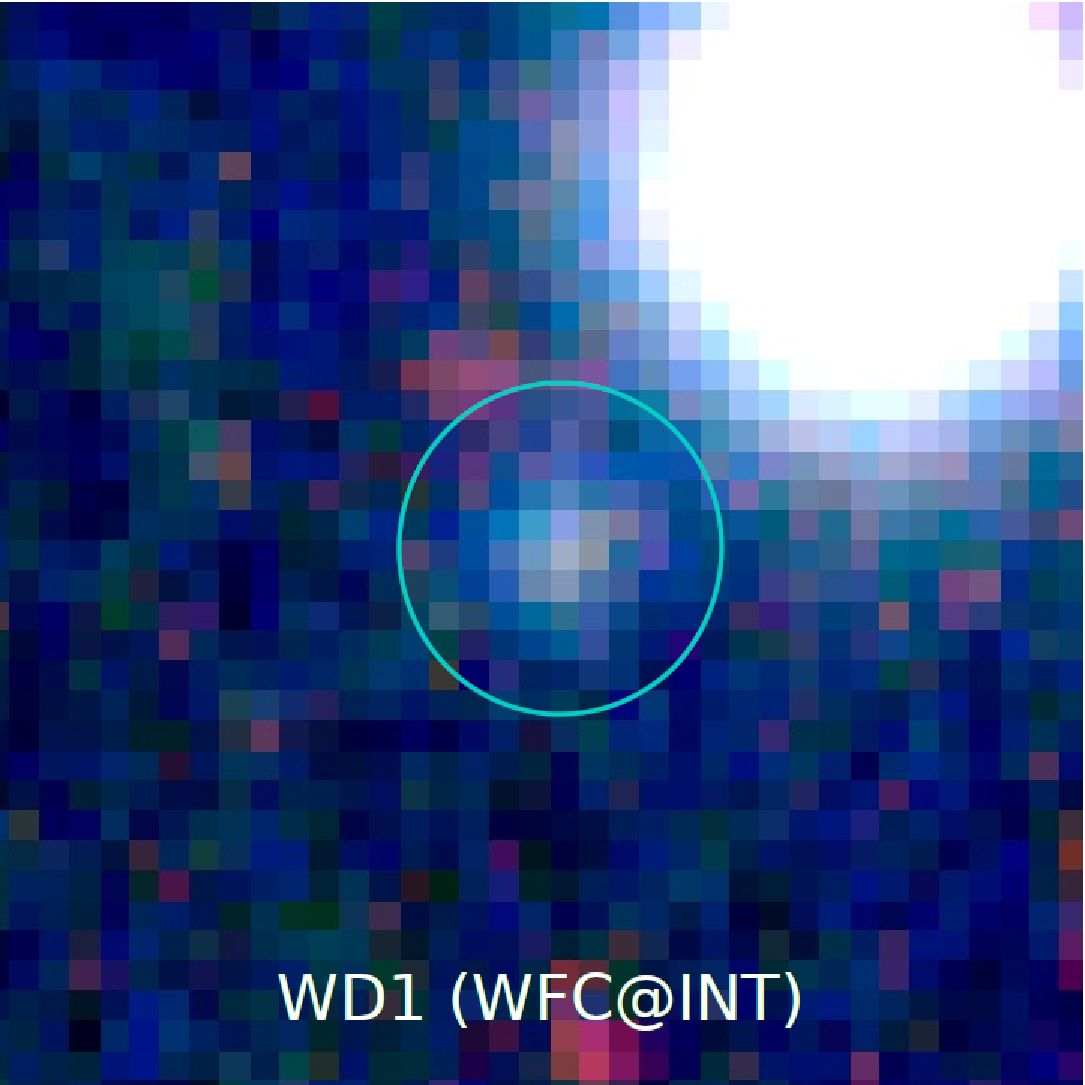} \\
       \includegraphics[width=0.125\textheight]{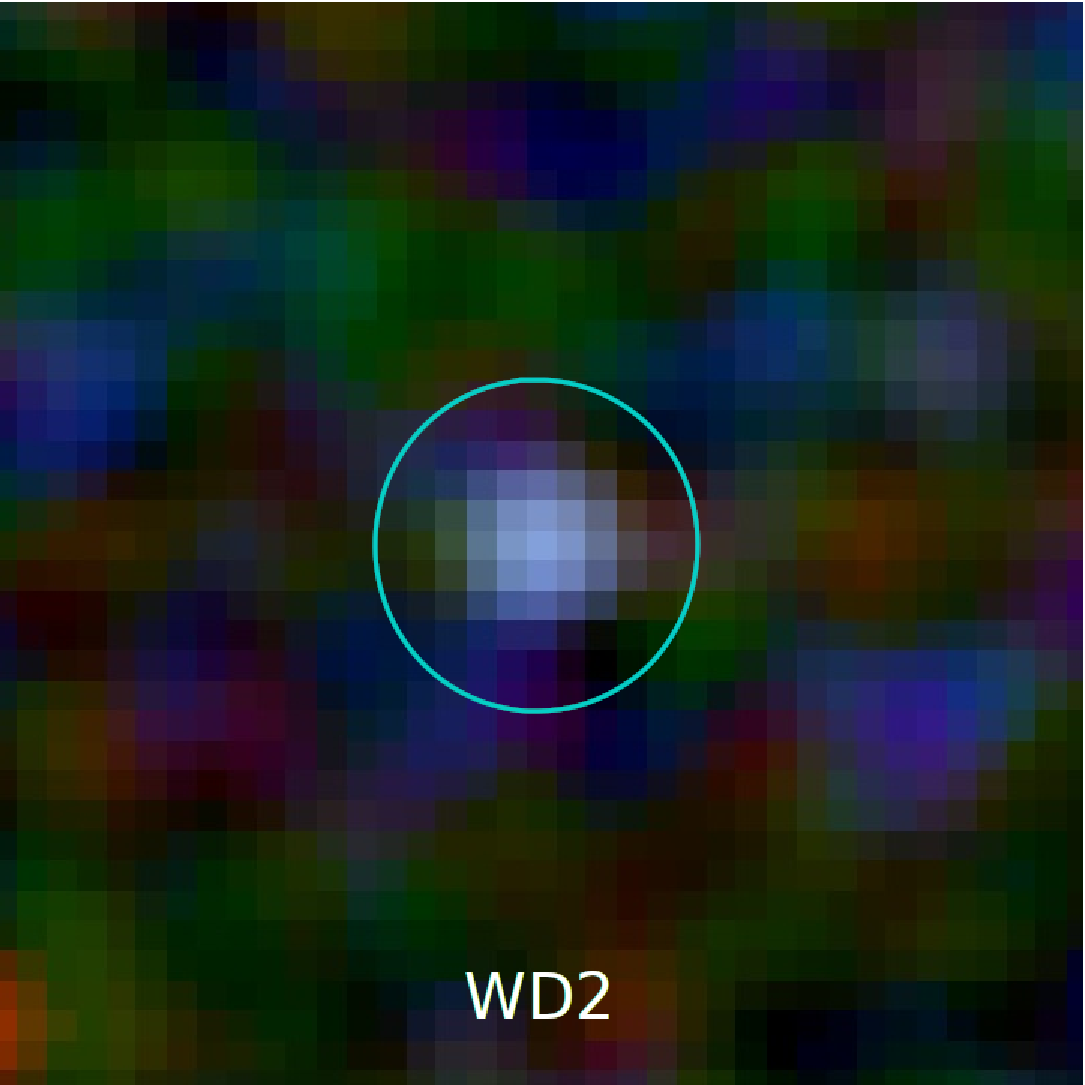}
       \includegraphics[width=0.125\textheight]{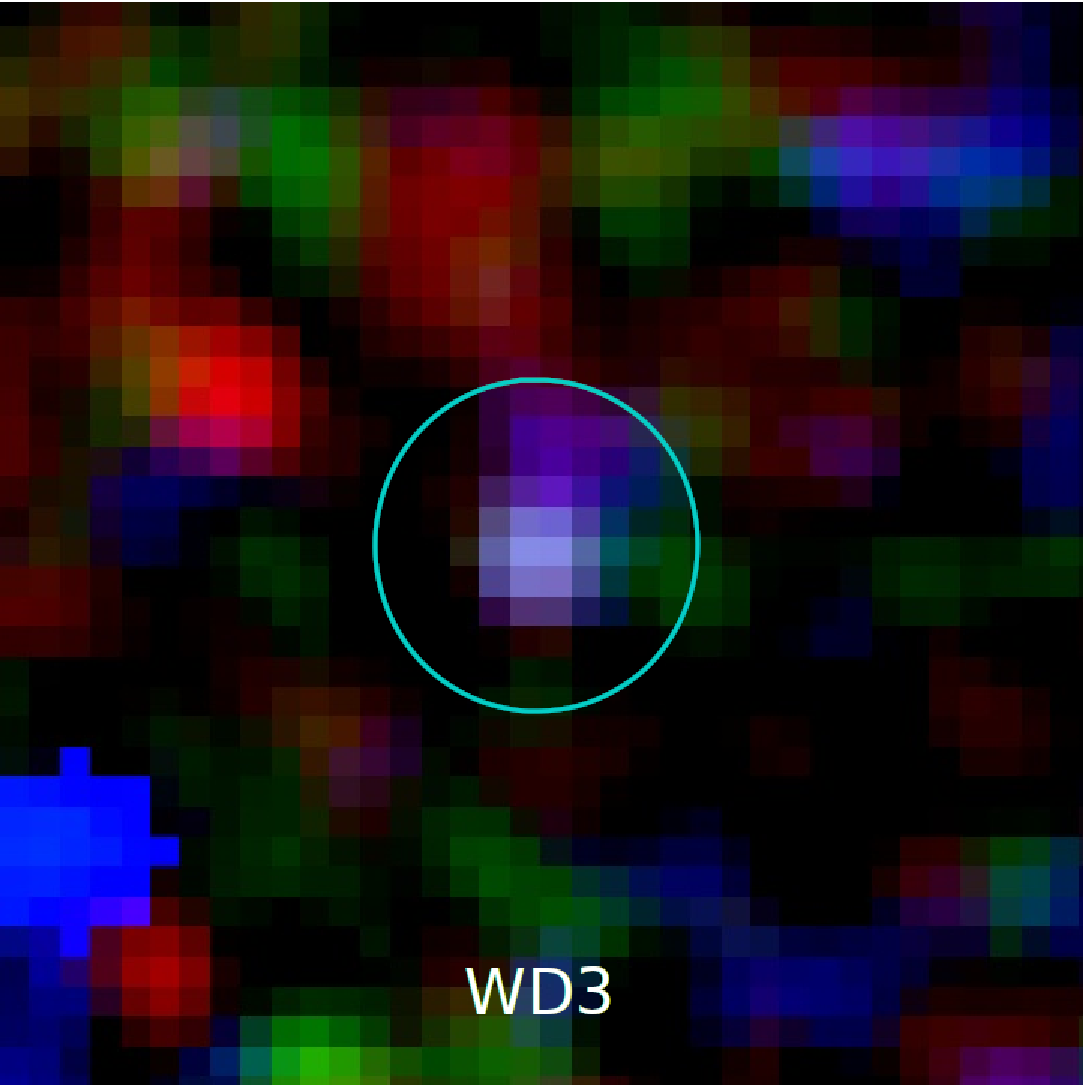} \\
       \includegraphics[width=0.125\textheight]{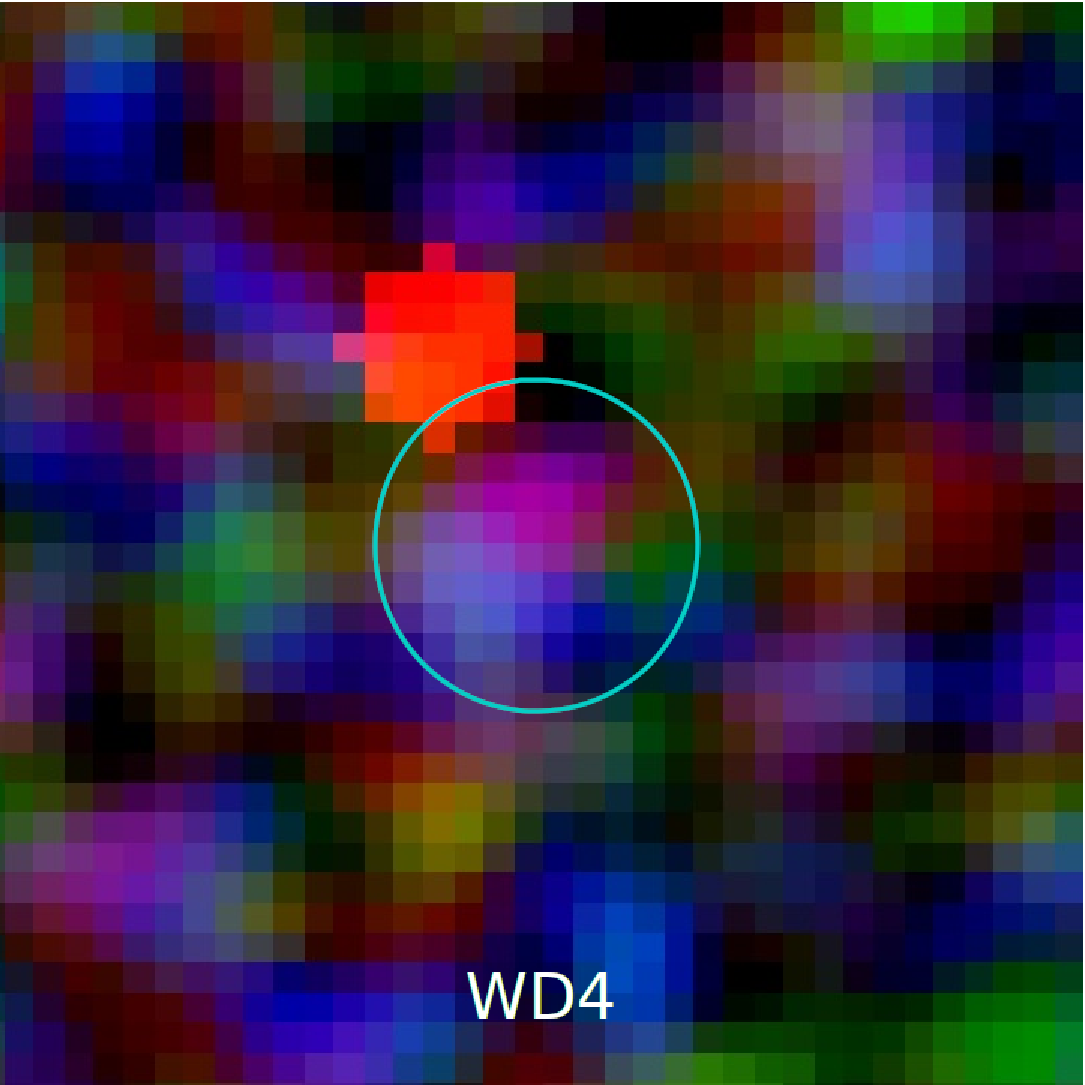}
       \includegraphics[width=0.125\textheight]{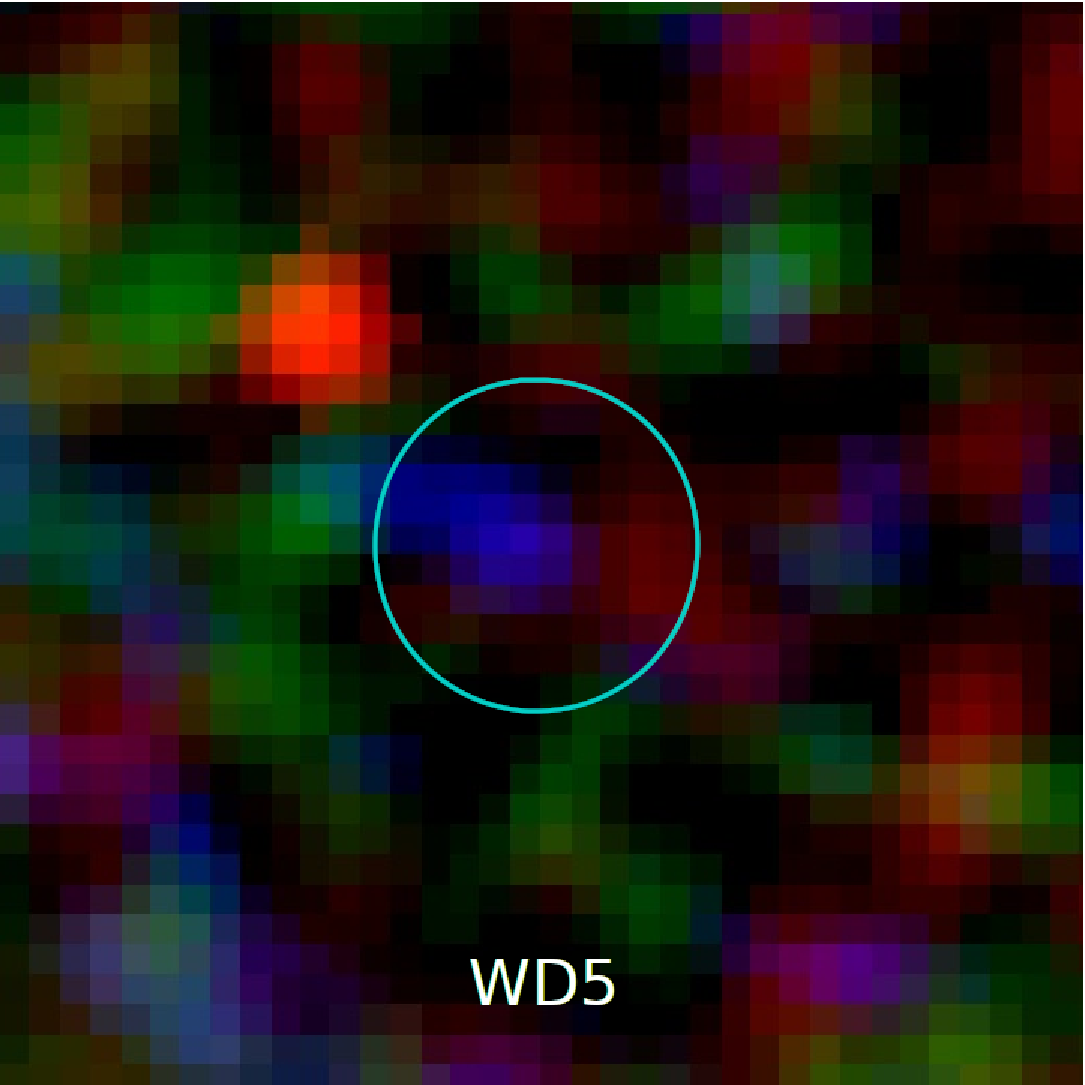} \\
     \end{subfigure} 
    \end{minipage}
   \caption{ Left panels show the $m_{\rm F090W}$ versus $(m_{\rm
       F070W}-m_{\rm F150W})$ and the $V$ versus $(V-I)$ CMDs of
     NGC\,2506 obtained with the NIRCam@{\it JWST} and WFC@INT data
     sets. In the leftmost panel the grey and black points are stars
     that passed the selection and membership criteria,
     respectively. Green and azure squares represent the sources that,
     based on their membership and/or position on the CMDs, may be
     considered candidate WDs. In red are reported the 2~Gyr WD
     isochrones; dotted lines represent the isochrone boundaries
     considering the errors on the distance modulus and reddening. The
     finding charts of the candidate WDs are shown in the right
     panels. The field of view of {\it JWST} finding charts is $1.2
     \times 1.2$~arcsec$^2$, while the WFC finding chart is $6.5
     \times 6.5$~arcsec$^2$. Sources WD1, WD2, and WD3 appear as
     point-like stars, while WD4 and WD5 should be noisy peaks.
      \label{fig:13}}
\end{figure*}

\begin{table*}
  \centering
  \caption{Candidate WDs identified in NGC\,2506}
  \resizebox{0.695\textwidth}{!}{\begin{tabular}{l c c c c c}
\hline
\hline
{\bf WD}  & {\bf $\alpha$} & {\bf $\delta$} & {\bf $m_{\rm F090W}$ } & {\bf $m_{\rm F070W}$ } & {\bf $m_{\rm F150W}$ } \\
 & (deg.) & (deg.) & & & \\
\hline
1  & 120.0166589 & $-$10.8069775 &  $22.95 \pm 0.05$ & $22.79 \pm 0.06$ & $23.16 \pm 0.01$ \\
2  & 120.0102662 & $-$10.8069270 &  $23.13 \pm 0.06$ & $22.94 \pm 0.03$ & $23.18 \pm 0.15$ \\
3  & 120.0571064 & $-$10.7940803 &  $23.98 \pm 0.02$ & $23.50 \pm 0.06$ & $23.75 \pm 0.13$ \\
4  & 120.0296222 & $-$10.7618422 &  $25.40 \pm 0.04$ & $25.11 \pm 0.03$ & $24.94 \pm 0.02$ \\
5  & 120.0466738 & $-$10.8187002 &  $25.52 \pm 0.01$ & $24.71 \pm 0.14$ & $24.88 \pm 0.10$ \\
\hline
\end{tabular}
}
  \label{tab:3}
\end{table*}

\subsection{Mass Functions and total mass}

We computed the luminosity functions (LFs) of MS stars counting stars
in magnitude intervals that contain an equal number of stars and using
both the Gaia and the {\it JWST} catalogue, and then we converted the
LFs to mass functions (MFs) by adopting the mass-luminosity relation
from the BaSTI-IAC isochrones.

As shown in panels (a) and (b) of Fig.~\ref{fig:12}, thanks to the
synergy between Gaia and {\it JWST} data, we are able to cover a mass
range from $\sim 0.1~M_{\odot}$ to $\sim 1.4~M_{\odot}$, encompassing
the very low mass stars and extending up to the main sequence
turn-off.

Panels (c) and (d) of Fig.~\ref{fig:12} show the MFs obtained with the
Gaia catalogue. They are also reported in Table~\ref{tab:4}. To
calculate the LFs, we accounted for the contribution of each star in a
given magnitude interval weighted by its MP to be a cluster member, as
already done in the previous sections. We focused on MS stars with
magnitudes between $G_{\rm RP}=14.50$ and $G_{\rm RP}=18.25$,
corresponding to a mass interval between $M=0.74~M_{\odot}$ and
$M=1.46~M_{\odot}$.  Initially, we considered the present day MF as
described by the equation:
\begin{equation}
\frac{{\rm d}N}{{\rm d}M} \propto M^{\alpha}
\end{equation}
which is a straight line 
in the logarithmic form:
\begin{equation}
\log{\left(\frac{{\rm d}N}{{\rm d}M}\right)} = \beta + {\alpha}\log{(M)}
\end{equation}
Panel (c) illustrates the MFs calculated in three radial bins
containing an equal number of MS stars. We fitted a straight line to
each MF within each radial bin using a weighted least square fit, with
the inverse square of the MF errors serving as the weights.  We
obtained three estimates of the present-day \textit{local} MF.  For
the inner region ($R<3.85$~arcmin), we derived a slope $\alpha=-0.99
\pm 0.25$; in the middle region ($3.85 \leq R<8.33$~arcmin) the slope
is $\alpha=-2.22 \pm 0.19$. Finally, for the outermost region ($8.33
\leq R<33.00$~arcmin) the slope measured is $\alpha=-2.80 \pm 0.24$.
We also calculated the slope of the MF using a single interval ($0.00
\leq R<33.00$~arcmin), and found $\alpha=-2.04 \pm 0.14$; i.e., a
proxy for the present-day \textit{global} mass function.  The change
in slope with radial distance from the centre is attributed to the
mass segregation effect resulting from the cluster dynamical
evolution, as previously noted by \citet{2013MNRAS.432.1672L} and
\citet{2019MNRAS.490.1383R}.\\

To determine the contribution to the cluster mass from stars with $M
\geq 0.74 M_\odot$, we utilised the MF shown in panel (d) of
Fig.\ref{fig:12}.  Our analysis yielded $M_{\rm cluster}[ 0.74
  M_{\odot} \leq M \leq 1.46 M_{\odot}] = 2077 \pm 67~M_{\odot}$.  We
also performed the same computation by using the three MFs shown in
panel (c) of Fig.~\ref{fig:12}: summing the single total mass
contributions in each radial bin, we obtained a mass $M_{\rm cluster}[
  0.74 M_{\odot} \leq M \leq 1.46 M_{\odot}] = 2054 \pm 69~M_{\odot}$.
This result aligns with the previously determined value, confirming
that mass segregation does not significantly impact the estimate of
the cluster mass obtained by integrating the global mass function
rather than using local mass functions.

We calculated the MFs in the area covered by the {\it JWST} dataset,
spanning approximately 20 arcmin$^2$ between 0.24 arcmin and 4.61
arcmin from the cluster centre.  The results are presented in
Table~\ref{tab:4} and in panel (e) of Fig.~\ref{fig:12}. First, we
calculated the MF for stars with masses between $0.74~M_{\odot}$ and
$1.46~M_{\odot}$ using Gaia~DR3 following the procedure previously
described (black points). For the {\it JWST} data, we adopted two
different approaches to compute the MFs. The first approach involved
the use of proper motions and was limited to stars with masses between
$0.37~M_{\odot}$ and $0.93~M_{\odot}$ (as ground-based INT images do
not reach magnitudes as faint as the {\it JWST} data, see
Sect.~\ref{sec:wfc}).  In this case, we followed a similar procedure
as for the Gaia data, weighting the MFs with MPs and correcting for
completeness (blue points in panel (e) of Fig.~\ref{fig:12}).
To extend the MF to lower masses, we performed a statistical
decontamination of the LFs to account for the contribution of field
stars. We selected stars in the Gaia catalogue located in the {\it
  JWST} region with magnitudes $16.5 \leq G_{\rm RP} \le 17.5$,
corresponding to a mass range between $0.84~M_{\odot}$ and
$1.03~M_{\odot}$ on the MS. Within this mass interval, which
corresponds to the magnitude range $15.67 \leq m_{\rm F277W} \le
16.42$, we calculated the weighted sum of the number of stars
considering the non-membership probabilities as weights.  Next, we
determined the mean number of expected field stars per unit F277W
magnitude by dividing the sum by 0.75 (the F277W magnitude range).  We
obtained a contamination factor of 22.8 stars/F277W magnitude.  We
subtracted this value appropriately from the number of stars counted
in each magnitude interval used to construct the LFs. The resulting MF
is represented by orange squares in panel (e) of
Fig.~\ref{fig:12}. Since the total MF does not conform to a single
power law, we fitted three straight lines to the ${\rm d}N/{\rm d}M$
vs. $M$ distribution in different mass intervals: for the mass range
$0.8 M_{\odot} < M \leq 1.4 M_{\odot}$, we obtained a power-law
exponent $\alpha = -1.74 \pm 0.22$; for masses $0.3 M_{\odot} < M \leq
0.8 M_{\odot}$, the best-fit was obtained for $\alpha = -0.82 \pm
0.17$; finally, in the low-mass regime $0.1 M_{\odot} \leq M \leq 0.3
M_{\odot}$ we measured $\alpha = -0.18 \pm 0.10$.

We computed the total mass for stars in the {\it JWST} region with
masses in the interval $0.10~M_{\odot} \leq M \leq 1.46~M_{\odot}$,
and obtained $M_{\rm cluster, JWST}[ 0.10 M_{\odot} \leq M \leq 1.46
  M_{\odot}] = 601 \pm 21~M_{\odot}$, of which $\sim 50~\%$ is due to
stars with masses between $0.74~M_{\odot}$ and $1.46~M_{\odot}$. We
found that the combined MF obtained from {\it JWST} and Gaia data can
be well described by the logistic function:
\begin{equation}
\log{\left(\frac{{\rm d}N}{{\rm d}M}\right)}= \frac{c}{1+a \exp{(-b \log{M})}}
\end{equation}
as shown in panel (f) of Fig.~\ref{fig:12}. The best-fitting parameters
were determined as $a = 0.231 \pm 0.020$, $b = -3.436 \pm 0.531$, and
$c=3.290 \pm 0.048$. By integrating this function, we obtained a total
mass $M_{\rm cluster, JWST}[ 0.10 M_{\odot} \leq M \leq 1.46
  M_{\odot}] = 614 \pm 77~M_{\odot}$, in agreement with the value
found by using three power laws.\\ 

We used this just derived information on the MFs within the
\textit{JWST} region to obtain an estimate for the present-day total
mass of the stellar cluster NGC\,2506 based only on stars with masses
ranging between 0.1 and 1.46~$M_{\odot}$.  In this calculation we
assume that the relative number of stars of different mass in the
entire cluster is the same as that observed within the annulus covered
by \textit{JWST}, i.e., within $0.24 \leq R \leq 4.61$~arcmin.  We
then calculated the total mass of the cluster in this annulus and
scaled it to the total region occupied by the cluster.  We obtained
$M_{\rm cluster, 0.24-4.61 arcmin} = M_{\rm cluster, JWST} \times
A({\rm annulus})/A({\rm JWST}) = 2021 \pm 150~M_{\odot} $ where
$A({\rm annulus})/A({\rm JWST})$ is the ratio between the area of the
annulus and the area covered by {\it JWST} observations. We then
calculated the cluster mass in an annulus between $4.61 \leq R \leq
33.00$~arcmin using stars with masses between 0.74 and
1.46~$M_{\odot}$ included in the Gaia catalogue; we found a mass of
$1223 \pm 47~M_{\odot}$, which is $\sim 59$~\% of the total mass
previously calculated for this mass range. Supposing that the mass
distribution is the same in the two radial bins, we calculated a total
mass equal to $\sim 5863~M_{\odot}$\footnote{In this calculation, we
excluded the central region $0.00 \leq R \leq 0.24$~arcmin, because in
the Gaia~catalogue only 7 cluster members are present, and it was not
possible to calculate the MF.  We considered the contribution of this
region negligible compared to the errors on the total mass.}. This
must be considered as a lower limit of the present-day total mass because it
does not take into account, i.e., of the evolved stars (giant stars,
white dwarfs, etc.), of stars with masses $<0.1~M_{\odot}$, and of the brown dwarfs.

\section{White dwarfs in NGC\,2506}
\label{sec:wd}
We investigated the presence of white dwarfs (WDs) in the {\it JWST}
CMDs of NGC\,2506 using the bluer, higher SNR filters, i.e. F070W,
F090W, and F150W, where WDs are expected to be more prominent.  The
results are presented in Fig.\ref{fig:13} and Table~\ref{tab:3}. The
leftmost panel shows the $m_{\rm F090W}$ versus $(m_{\rm F070W}-m_{\rm
  F150W})$ CMD of the stars that passed the selection and membership
criteria in grey and black, respectively. The isochrone of WDs with an
age of 2~Gyr is depicted in red. This WD isochrone has been calculated
as described in \citet{griggiowd} using the BaSTI-IAC WD cooling
models by \citet{bastiwd} for progenitors with initial
[M/H]=$-$0.40. We obtain essentially the same isochrone if we use
cooling models from progenitors with [M/H]=$-$0.20, the other
metallicity in the WD BaSTI-IAC database close to the cluster [M/H].
The green square shows the only WD with a high-membership probability
(92.1~\%), as consistently located along the WD cooling sequence.
According to the initial-final mass relation adopted in the
calculation of the isochrone \citep[][]{ifmr} the mass of this WD is
$\sim$0.62$M_{\odot}$.

This same star (WD1, $V=22.74 \pm 0.08$, $I=23.07 \pm 0.03$) is also
displayed in the $V$ versus $(V-I)$ CMD obtained with the WFC@INT data
set, where it also closely matches the WD isochrone within the
measurement errors. The azure squares in the leftmost panel represent
other sources near the WD isochrone but without proper motion
measurements, which should be considered as candidate WDs.  A suitable
astrometric second epoch appears as the most efficient way to confirm
as members these WD candidates.  We visually examined these five
sources using three-colour stacked images obtained with the same
filters employed for the CMDs. The results are presented in the right
panels: candidate WDs 1, 2, and 3 appear as clear point-like sources
in both the {\it JWST} and WFC@INT images. On the other hand, WD4 and
WD5 likely correspond to noisy peaks that passed the selection
criteria. However, with deeper data collected by either \textit{HST}
or {\it JWST} in the future, this ambiguity may be resolved.

\noindent

\section{Summary}
\label{sec:sum}
In this work we have exploited {\it JWST} non-proprietary calibration
data to: (i) derive accurate effective PSFs for ten filters (8 wide + 2
medium), spanning a wavelength interval from 0.7~$\mu$m to 4.5~$\mu$m;
(ii) extract high-precision photometry and astrometry from "shallow" NIRCam images
for stars located in the region that encompasses a portion of the 2
Gyr open cluster NGC\,2506; (iii) calculate the proper motions for MS
stars with masses $\gtrsim 0.3~M_{\odot}$ by adopting ground-based
data collected with the INT a taking advantage of the large temporal
baseline ($\sim 18.8$~years) between {\it JWST} and INT data; (iv)
carry-out an in-depth analysis of the cluster properties by leveraging
the synergy between {\it JWST} data and the Gaia~DR3 catalogue.

We have calculated the radial stellar density profile by using the
Gaia~DR3 catalogue, fitted with a King profile. From the fitting, we
derive the central stellar density ($k_0=28.1 \pm
0.9$~stars/arcmin$^2$), the core radius ($r_{\rm c}=2.60 \pm
0.05$~arcmin), and the tidal radius ($r_{\rm t}=33.0 \pm 4.3$~arcmin).

From the combination of {\it JWST} and Gaia~DR3 data, we calculated
the fraction of MS binaries with mass ratio $q \geq 0.6$, equal to
$\sim 23$~\%. This synergy allowed us to extend the study of the MS
binary fraction down to $\sim 0.4~M_{\odot}$, well below the Gaia
limit ($\sim 0.8~M_{\odot}$).  Our findings reveal no dependence of
the binary fraction on the primary mass within a mass range between
$0.44 M_{\odot}$ and $1.13 M_{\odot}$. However, we observed a hint of
a radial trend in the radial distribution of the MS binary fraction
between the cluster centre and approximately $2 r_{\rm c}$.

Similarly, by leveraging the synergy between {\it JWST} and Gaia~DR3,
we computed MFs of MS stars within the mass interval from $\sim 0.10
M_{\odot}$ to $\sim 1.45 M_{\odot}$. First, we examined the MFs using
solely Gaia~DR3 data. Our analysis confirmed the influence of mass
segregation on the MFs calculated in various radial bins for masses
ranging from $\sim 0.7~M_{\odot}$ to $\sim 1.4~M_{\odot}$. By
integrating these mass functions, we determined that the total mass of
stars within the range from $0.74~ M_{\odot}$ to $1.46~M_{\odot}$ is
$\sim 2065~M_{\odot}$.  We then focused on the region covered by {\it
  JWST} data ($0.24 \leq R \leq 4.61$ arcmin). We found that the MF
for stars with masses between $\sim 0.1~M_{\odot}$ and $\sim
1.4~M_{\odot}$ is well represented by a logistic function. By
calculating the total mass of stars within this region ($\sim 600
M_{\odot}$) and assuming a homogeneous mass distribution, we estimated
the total mass within $R \leq 4.61$~arcmin to be $\sim
2021~M_{\odot}$. We used this number to put a lower limit on the
total mass of the cluster, which is estimated to be about
$6000~M_{\odot}$.

Finally, using the bluest {\it JWST} available filters, we identified
five candidate white dwarfs. Among them, WD1 stands out as the
strongest candidate due to its relatively high brightness, the high
membership probability and alignment with the theoretical 2 Gyr white
dwarf cooling sequence (also on the ground-based CMD). A preliminary
estimate suggests that the mass of WD1 is $\sim
0.62~M_{\odot}$. However, to resolve the ambiguity surrounding these
candidates, it will be necessary to acquire new, deeper data using
instruments such as {\it HST} or {\it JWST}.

As a byproduct of this work, we release the derived ePSFs and the {\it
  JWST} catalogues and atlases of NGC\,2506 (see Appendix~\ref{app:1}).

This study has demonstrated the potential of using {\it JWST}'s
publicly available calibration images in combination with the Gaia
catalogue, to conduct a pioneering multi-band analysis of stellar
populations in open clusters within the IR wavelength range. With a
minimum effort of observing time, we were able to probe the faint end
of the MS in this peculiar, old, and distant open cluster. Future
NIRCam deeper observations of different open clusters, spanning
multiple years, will enable us to easily explore their brown dwarf
sequence, study the mass functions and evaporation effects of low-mass
stars ($<0.2~M_{\odot}$), the internal kinematic via proper-motions,
as well as investigate their white dwarf cooling sequences, advancing
our understanding of stellar populations in star clusters and shedding
light on the diverse phenomena within these systems.

%#######################################################
\section*{Acknowledgements}

The authors warmly thank Dr F. Van Leeuwen for the careful
  reading and suggestions that improved the quality of our paper.
DN, LRB, and MG acknowledge support by MIUR under PRIN programme
\#2017Z2HSMF and by PRIN-INAF 2019 under programme \#10-Bedin.  This
work has made use of data from Pan-STARRS1 Surveys. The Pan-STARRS1
Surveys (PS1) and the PS1 public science archive have been made
possible through contributions by the Institute for Astronomy, the
University of Hawaii, the Pan-STARRS Project Office, the Max-Planck
Society and its participating institutes, the Max Planck Institute for
Astronomy, Heidelberg and the Max Planck Institute for
Extraterrestrial Physics, Garching, The Johns Hopkins University,
Durham University, the University of Edinburgh, the Queen's University
Belfast, the Harvard-Smithsonian Center for Astrophysics, the Las
Cumbres Observatory Global Telescope Network Incorporated, the
National Central University of Taiwan, the Space Telescope Science
Institute, the National Aeronautics and Space Administration under
Grant No. NNX08AR22G issued through the Planetary Science Division of
the NASA Science Mission Directorate, the National Science Foundation
Grant No. AST-1238877, the University of Maryland, Eotvos Lorand
University (ELTE), the Los Alamos National Laboratory, and the Gordon
and Betty Moore Foundation.

This work has made use of data from the European Space Agency (ESA)
mission {\it Gaia} (\url{https://www.cosmos.esa.int/gaia}), processed
by the {\it Gaia} Data Processing and Analysis Consortium (DPAC,
\url{https://www.cosmos.esa.int/web/gaia/dpac/consortium}). Funding
for the DPAC has been provided by national institutions, in particular
the institutions participating in the {\it Gaia} Multilateral
Agreement.

\section*{Data Availability}
The data underlying this article are publicly available in the
Mikulski Archive for Space Telescopes at
\url{https://mast.stsci.edu/}. The catalogues underlying this work
are available in the online supplementary material of the article.

%%%%%%%%%%%%%%%%%%%% REFERENCES %%%%%%%%%%%%%%%%%%

% The best way to enter references is to use BibTeX:
\bibliographystyle{mnras}
\bibliography{biblio}

\appendix
\section{PSF variations across the detectors}
\label{app:2}
\begin{figure*}
\includegraphics[width=0.9\textwidth]{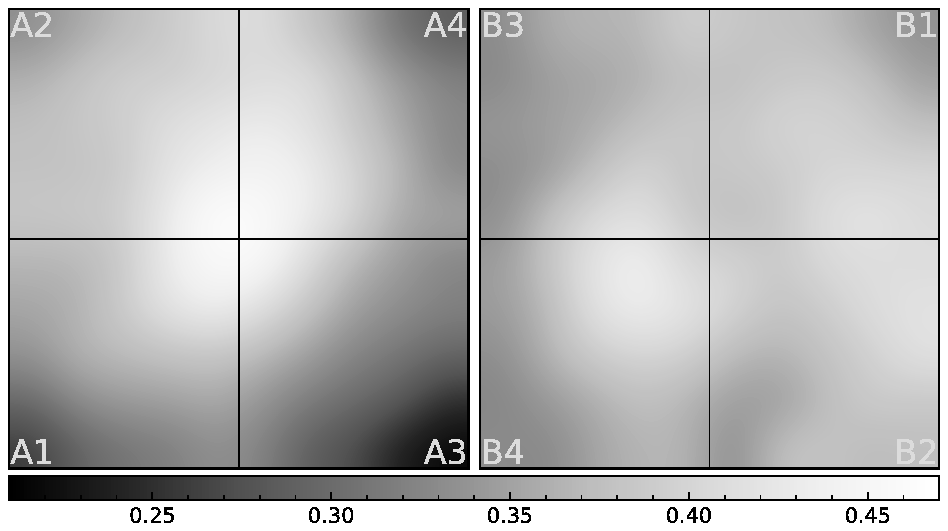}
\caption{Variation across the detector of the peak of the PSF in F070W filter.}\label{fig:b1}
\end{figure*}
Given the great level of accuracy of our here-derived
second-generation PSFs, significantly improved (if compared with the
PSFs from Paper\,\citetalias{2022MNRAS.517..484N}) thanks to the
available geometric distortion corrections (derived only in
Paper\,\citetalias{2023AN....34430006G}), we can now provide the
community with an independent determination of the average spatial
properties of the PSFs.  Indeed, it would be particularly useful to
know precisely where to place a target in order to obtain the highest
possible angular resolution with NIRCam@{\it JWST} observations.
Given the dependency of the angular resolution on the wavelength for a
diffraction-limited `optical' (actually, IR) system, the bluest
filters are those expected to show the maxima spatial variations, as
they would amplify relative differences.  Furthermore, the
under-sampled nature of NIRCam PSFs of bluest filters makes the PSFs
of the filters F070W/F090W/F115W the hardest PSFs to solve for.

In Fig.\,\ref{fig:b1}, we show the spatial determination of PSFs's
peaks in the most undersampled filter F070W, obtained by interpolating
the 5$\times$5 perturbed PSF arrays of all the images.  PSFs values
are normalised to 1 within an area of 5.25$\times$5.25 physical
pixels. So a value of 0.47 meaning that the central pixels, of a
source centred at the centre of a pixels, contain 47\% of the
normalised flux.  Figure\,\ref{fig:b1} reveals that Module A (on the
left) has a sharpest 'sweet spot' in a region between the four
detectors, where targets that need the highest angular resolution
(FWHM$\sim$35\,mas, containing 47\% of the light in its central pixel)
should be placed.  Also Module B shows a slightly less-peaked sweet
spot than Module A, and relatively off-centre, mainly within B4.

These differences between PSFs at different spatial position within
NIRCam field of view, become less and less important moving to redder
wavelengths.

\section{Electronic Material}
\label{app:1}
The catalogues of NGC\,2506 extracted in this work will be released as
supporting material to this paper. We will release two catalogues, one
for each field of view covered by Module A and B, that contain
information on the position of the stars, the VEGAmag calibrated
magnitudes in the 10 {\it JWST} filters, the quality parameters
described in Sect.~\ref{sec:catalog}, the quality flag of the
selections described in Sect.~\ref{sec:catalog}; the proper motions
and the membership probabilities, and the completeness associated to
each star. We will also make publicly available the ePSFs and the
stacked images in each filter derived in this work at the website:
\texttt{https://web.oapd.inaf.it/bedin/files/\-PAPERs\_eMATERIALs/JWST/}. At
the same website we will upload the NGC\,2506 catalogues.

\label{lastpage}
\end{document}